\documentclass[a4paper,english,journal]{IEEEtran}
\usepackage[T1]{fontenc}
\usepackage[latin9]{inputenc}
\usepackage{color}
\usepackage{textcomp}
\usepackage{amsthm}
\usepackage{amsmath}
\usepackage{amssymb}
\usepackage{graphicx}

\makeatletter

\providecommand{\tabularnewline}{\\}

\theoremstyle{plain}
\newtheorem{thm}{\protect\theoremname}
\theoremstyle{definition}
\newtheorem{defn}[thm]{\protect\definitionname}

\usepackage[font={small}]{caption}\usepackage{changebar}\usepackage[noadjust]{cite}\usepackage{array}\usepackage[ruled,lined,linesnumbered,longend]{algorithm2e}
\usepackage{url}
\usepackage{babel}
\providecommand{\definitionname}{Definition}
\providecommand{\theoremname}{Theorem}

\usepackage{babel}
\providecommand{\definitionname}{Definition}
\providecommand{\theoremname}{Theorem}

\usepackage{babel}
\providecommand{\definitionname}{Definition}
\providecommand{\theoremname}{Theorem}

\@ifundefined{showcaptionsetup}{}{%
 \PassOptionsToPackage{caption=false}{subfig}}
\usepackage{subfig}
\makeatother

\usepackage{babel}
\providecommand{\definitionname}{Definition}
\providecommand{\theoremname}{Theorem}

\begin{document}

\title{Spectral and Energy Spectral Efficiency Optimization of Joint Transmit
and Receive Beamforming Based Multi-Relay MIMO-OFDMA Cellular Networks}

\author{Kent~Tsz~Kan~Cheung, Shaoshi~Yang,~\IEEEmembership{Member,~IEEE},
and Lajos~Hanzo,~\IEEEmembership{Fellow,~IEEE}%
\thanks{This research has been funded by the Industrial Companies who are
Members of the Mobile VCE, with additional financial support from
the UK Government's Engineering \& Physical Sciences Research Council~(EPSRC).
The financial support of the Research Councils UK~(RCUK) under the
India-UK Advanced Technology Center~(IU-ATC), of the EU under the
auspices of the Concerto project, and of the European Research Council's
Senior Research Fellow Grant is also gratefully acknowledged.

The authors are with the School of Electronics and Computer Science,
University of Southampton, Southampton, SO17 1BJ, UK (e-mail: \{ktkc106,shaoshi.yang,lh\}@ecs.soton.ac.uk).%
}\vspace{-8mm}
 }
\maketitle
\begin{abstract}
We first conceive a novel transmission protocol for a multi-relay
multiple-input--multiple-output orthogonal frequency-division multiple-access~(MIMO-OFDMA)
cellular network based on joint transmit and receive beamforming.
We then address the associated network-wide spectral efficiency~(SE)
and energy spectral efficiency~(ESE) optimization problems. More
specifically, the network's MIMO channels are mathematically decomposed
into several effective multiple-input--single-output~(MISO) channels,
which are essentially spatially multiplexed for transmission. Hence,
these effective MISO channels are
referred to as spatial multiplexing components~(SMCs). For
the sake of improving the SE/ESE performance attained, the SMCs are
grouped using a pair of proposed grouping algorithms.
The first is optimal in the sense that it exhaustively evaluates all
the possible combinations of SMCs satisfying both the semi-orthogonality
criterion and other relevant system constraints, whereas the second
is a lower-complexity alternative. Corresponding to each of the two
grouping algorithms, the pair of SE and ESE maximization problems
are formulated, thus the optimal SMC groups and optimal power control
variables can be obtained for each subcarrier block.
These optimization problems are proven to be concave, and the dual
decomposition approach is employed for obtaining their solutions.
Relying on these optimization solutions, the impact
of various system parameters on both the attainable SE and ESE is
characterized. In particular, we
demonstrate that under certain conditions the lower-complexity SMC
grouping algorithm achieves $90\%$ of the SE/ESE attained by the
exhaustive-search based optimal grouping algorithm, while imposing
as little as $3.5\%$ of the latter scheme's computational complexity.\end{abstract}

\section{Introduction\label{sec:Intro}}

\IEEEPARstart{R}{ecent} wireless mobile broadband standards optionally
employ relay nodes~(RNs) and multiple-input--multiple-output orthogonal
frequency-division multiple-access~(MIMO-OFDMA) systems~\cite{Salem2010,Hanzo2010}
for supporting the ever-growing wireless capacity demands. These systems
benefit from a capacity gain increasing roughly linearly both with
the number of available OFDMA subcarriers~(each having the same bandwidth)
as well as with the minimum of the number of transmit antennas~(TAs)
and receive antennas~(RAs). However, this capacity-oriented approach
conflicts with the increasing need to reduce the system's carbon footprint~\cite{Han2011}
as increasing the number of radio frequency~(RF) chains and subcarriers
will incur additional energy costs. \emph{In light of these discussions,
the goal of this paper is to formally
optimize the spectral efficiency~(SE) or energy spectral efficiency~(ESE)
of the downlink~(DL) in a multi-relay MIMO-OFDMA cellular system
by intelligently allocating the
available power and frequency resources and employing joint transmit
and receive beamforming~(BF).}

It is widely acknowledged that under the idealized simplifying condition
of having perfect channel state information~(CSI) at the transmitter,
the DL or broadcast channel~(BC) capacity~\cite{Caire2003,Vishwanath2003}
may be approached with the aid of dirty paper coding~(DPC)~\cite{Costa1983}.
However, the practical implementation of DPC
is hampered by its excessive algorithmic complexity upon increasing
the number of users. On the other hand, BF is an attractive suboptimal
strategy for allowing multiple users to share the BC while resulting
in reduced multi-user interference~(MUI). A low-complexity transmit-BF
technique is the zero-forcing based BF~(ZFBF), which can asymptotically
achieve the BC capacity as the number of users tends to infinity~\cite{Yoo2006}.
Furthermore, ZFBF may be readily applied to a system with multiple-antenna
receivers through the use of the singular value decomposition~(SVD).
As a result, the associated MIMO channels may be
mathematically decomposed into several \emph{effective} multiple-input--single-output~(MISO)
channels, which are termed spatial multiplexing components~(SMCs)%
\footnote{Note that these effective MISO channels are different from the physical
MISO channels directly composing the physical MIMO channel. For brevity,
we coin the term SMC to emphasize that these effective MISOs will
be used for the purpose of spatial multiplexing. A more in-depth discussion
regarding the concept of SMCs will be provided in Section~\ref{sec:TxProt}.%
} in this work. Furthermore, in~\cite{UlHassan2009}, these SMCs are
specifically grouped so that the optimal grouping as well as the optimal
allocation of the power may be found on each subcarrier
block using convex optimization. In contrast to the
channel-diagonalization methods of~\cite{Raleigh1998,Wong2003,Ho2009},
the ZFBF approach does not enforce any specific relationship between
the total numbers of TAs and RAs. Therefore, ZFBF is more suitable
for practical systems, since the number of TAs at the BS is typically
much lower than the total number of RAs of all the active user equipments~(UEs).
Compared to the random beamforming methods, such as that
of~\cite{Sharif2005}, ZFBF is capable of completely
avoiding the interference, allowing
us to formulate our SE/ESE maximization~(SEM/ESEM) problems as convex
optimization problems. Due to its desirable performance versus complexity
trade-off, in this paper we employ
ZFBF in the context of multi-relay aided MIMO-OFDMA systems, where
the direct link between the base station~(BS) and the UE may be exploited
in conjunction with the relaying link for further improving the system's
performance.

We formally define the ESE as a counterpart of the area spectral efficiency~(ASE)~\cite{Goldsmith2005},
where the latter has the units of $\left[\mbox{bits/sec/Hz/m}^{2}\right]$,
while the former is measured in $\left[\mbox{bits/sec/Hz/Joule}\right]$.
The ESE metric has been justified, for example, in~\cite{Ng2012a,Miao2010,Miao2012,Cheung2013}.
However, these contributions did not consider resource allocation
in the context of a MIMO system, and only~\cite{Cheung2013} incorporated
relaying. On the other hand, although there are numerous contributions
on optimal resource allocation in MIMO systems, they typically only
focused on either the SEM~(equivalently, the sum-rate maximization)
or the power minimization~\cite{UlHassan2009,Shen2006,Yu2007,Lo2007,Ng2012c}.
For example, the authors of~\cite{UlHassan2009}
applied BF to a DL cellular system and aimed for minimizing the resultant
total transmission power, while simultaneously satisfying the per-user
rate requirements. The authors of~\cite{Yu2007} instead choose to
minimize the per-antenna transmission powers, while satisfying both
the maximum per-antenna power constraints as well as the per-user
signal-to-noise-plus-interference~(SINR) requirements. Although there
exists some literature studying the ESE of relay-aided MIMO systems~\cite{Brante2013,Zappone2014},
these contributions typically focus their attention on a simple three-node
network consisting of the source, the destination and a single RN.

To summarize, there is a paucity of literature on the convex optimization
approach to the ESEM problem associated with both resource allocation
and joint transmit/receive beamforming in the context of multi-user
multi-relay MIMO-OFDMA systems. Additionally, the Charnes-Cooper transformation~\cite{Avriel1988}
is employed in this paper for solving the associated ESEM problem,
in contrast to the scalarization approach~\cite{Devarajan2012} that
requires the weighting of multiple objectives. On the other hand,
the Dinkelbach's method~\cite{Dinkelbach1967,Ng2012a,Ng2012,Cheung2013}
is avoided as it would require solving
a series of parametric convex problems, rather than the resultant
single convex problem of the Charnes-Cooper transformation. Although
the latter approach does impose an additional linear constraint on
the problem, in our experience,
this only marginally increases the complexity of the solution algorithm.
The authors of~\cite{Isheden2012} employed the Charnes-Cooper variable
transformation for the ESEM of a simple point-to-point link. However,
as far as we are aware, the Charnes-Cooper transformation has rarely
been used in other contexts for solving the ESEM problem.

Let us now summarize the above discussions and provide a concise list
of the novel contributions of this paper: 
\begin{itemize}
\item \textbf{We consider a generalized multi-user multi-relay assisted
MIMO-OFDMA cellular system model for the SEM/ESEM problems.} To provide
some justification, this system model accounts for both the direct
links between the BS and the UEs, as well as the relaying links employing
the decode-and-forward~(DF) relaying protocol~\cite{Laneman2004}.
This system model is unlike that of~\cite{Yoo2006,UlHassan2009},
which did not consider relaying, and it is also
distinct from that of~\cite{Brante2013,Zappone2014},
which only consider a single RN and a single UE. Additionally,
we dispense with the constraint that the number of antennas at the
BS needs to be greater or equal to the sum of the number of antennas
at the UEs, which was assumed in~\cite{Raleigh1998,Wong2003,Ho2009}.
Furthermore, this system model is built upon our previous work~\cite{Cheung2013}
as the network elements may now be equipped with an arbitrary number
of antennas for improving the system's SE or ESE performance. 
\item \textbf{A sophisticated novel transmission protocol is proposed for
improving the system's SE/ESE performance.} Since the multi-relay
MIMO-OFDMA system model considered
has not been studied in the context of the SEM/ESEM problems before,
we develop a novel transmission protocol that exploits spatial multiplexing
in both transmission phases while allowing both the direct and relaying
links to be simultaneously active. Although this
protocol does not benefit from a higher spatial degree of freedom
than that of the conventional half-duplex relay based cooperative
system, we glean more flexibility in choosing the best group of channels
for each transmission phase, which leads to additional selection diversity.
As a result, the achievable SE/ESE performance may be improved. Again,
this protocol is distinct from that presented in~\cite{Yoo2006,UlHassan2009},
since relaying is not considered in those works. Another benefit is
that since spatial multiplexing is employed in conjunction with OFDMA,
multiple data streams may be served using \emph{the same} subcarrier
block, while the transmit ZFBF is employed for avoiding the interference.
Furthermore, the receive-BF matrices are designed with the aim of
generating a number of SMCs that
may be grouped for the purpose of increasing the attainable spatial
multiplexing gain. 
\item \textbf{Two SMC grouping algorithms are proposed.} To elaborate, we
present a pair of novel algorithms for grouping the SMCs transmissions.
The possibility of relayed transmissions means that we have to partition
each transmission period into two halves, one consisting of BS-to-UE
and BS-to-RN links, and the other consisting of additional BS-to-UE
as well as RN-to-UE links. As a result, the SMC-pairs
of the two-hop relaying links are incomparable to the SMCs of the
direct links in either the first or the second transmission phases.
This is because, firstly the RNs are subject to their individual maximum
transmission power constraints, and secondly they employ the DF protocol,
which means that the information conveyed on the RN-to-UE link cannot
be more than that conveyed on the BS-to-RN link. These challenging
issues are resolved by the proposed grouping algorithms. The first
grouping algorithm is optimal in the sense that
it is based on exhaustive search over all the SMC
groupings that satisfy the semi-orthogonality criterion, while the
second algorithm constitutes a lower-complexity alternative. This
complexity-reduction is achieved by a multi-stage
SMC group construction process. In each stage, we firstly compute
the orthogonal components with respect to the vectors contained in
the tentative SMC group to be constructed using all the residual legitimate
SMC vectors, and then insert the particular SMC vector that results
in the orthogonal component having the highest norm into the SMC group
to be constructed. In principle, this method is similar
to that of~\cite{Yoo2006,UlHassan2009}, but it has been appropriately
adapted for the multi-relay cellular network considered under the
above-mentioned particular constraints. 
\item \textbf{The problems of choosing the SE- or ESE-optimal SMC groupings
and their associated power control values are formulated and solved
using convex optimization.} In contrast to~\cite{UlHassan2009,Shen2006,Yu2007,Lo2007,Ng2012c},
the crucial objective of maximizing the ESE metric is employed, as
motivated above. On the other hand, in contrast to~\cite{Ng2012a,Miao2010,Miao2012},
we consider a system that allows for \emph{simultaneous} direct and
relayed transmissions for the sake of increasing the attainable spatial
multiplexing gain. Although there exist other methods of solving this
ESEM problem~\cite{Devarajan2012,Ng2012a,Ng2012,Cheung2013}, we
employ the Charnes-Cooper transformation~\cite{Avriel1988} for obtaining
the maximum ESE solution, as it exhibits a reduced complexity from
having to solve only a single convex optimization problem. 
\end{itemize}
The rest of this paper is organized as follows. Section~\ref{sec:SysModel}
describes the multi-relay MIMO-OFDMA cellular network considered,
while Section~\ref{sec:TxProt} characterizes our novel transmission
protocol that allows for both direct and relaying links to be simultaneously
and continuously activated. In Section~\ref{sec:SGAlgor}, we elaborate
on the aforementioned SMC grouping algorithms conceived for forming
the sets of possible SMC transmission groups. The
issue of finding the optimal SMC transmission groups
and the optimal power control variables is then formulated as an optimization
problem in Section~\ref{sec:ProbForm}, which is then solved by using
a number of variable transformations and relaxations. The performance
of both our SMC grouping algorithms
and of the SEM/ESEM solvers are characterized in Section~\ref{sec:ResDis}.
Finally, we present our conclusions and future research ideas in Section~\ref{sec:Conc}.

\section{System model\label{sec:SysModel}}

\begin{figure}
\begin{centering}
\includegraphics[scale=0.54]{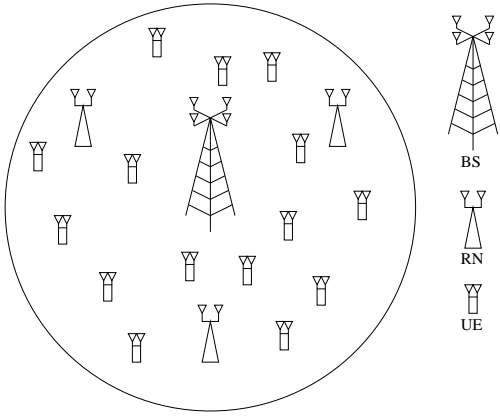} 
\par\end{centering}

\protect\caption{An example of a multi-relay MIMO-OFDMA cellular network, consisting
of a single BS at the cell-center, 3 RNs and 15 UEs.}

\label{fig:cellular} 
\end{figure}

We focus our attention on the DL of a multi-relay MIMO-OFDMA cellular
network, as shown in Fig.~\ref{fig:cellular}. The BS, $M$ DF-assisted
RNs and $K$ UEs are each equipped with $N_{B}$, $N_{R}$ and $N_{U}$
antennas, respectively. The cellular system has access to $N$ subcarrier
blocks, each encompassing $W$ Hertz of wireless bandwidth. The subcarrier
blocks considered here are similar to the resource blocks in the LTE-nomenclature~\cite{3GPP_PL}.
The BS is located at the cell-center, while the RNs are each located
at a fixed distance from the BS and are evenly spaced around it. The
ratio of the distance between the BS and RNs to the cell radius is
denoted by $D_{r}$. On the other hand, the UEs are uniformly distributed
in the cell. The BS coordinates and synchronizes its own transmissions
with that of the RNs, which employ the DF protocol and
thus avoids the problem of noise amplification. As it will be shown in Section~\ref{sub:OptPrimal},
this strategy results in a simple algorithm for finding the
optimal power control variables.

For the subcarrier block $n\in\left\{ 1,\cdots,N\right\} $, let us
define the complex-valued wireless channel matrices between the BS
and UE $k\in\left\{ 1,\cdots,K\right\} $, between the BS and RN $m\in\left\{ 1,\cdots,M\right\} $,
and between RN $m$ and UE $k$ as $\mathbf{H}_{n,k}^{BU}\in\mathbb{C}^{N_{U}\times N_{B}}$,
$\mathbf{H}_{n,m}^{BR}\in\mathbb{C}^{N_{R}\times N_{B}}$ and $\mathbf{H}_{n,m,k}^{RU}\in\mathbb{C}^{N_{U}\times N_{R}}$,
respectively. These complex-valued channel matrices account for both
the frequency-flat Rayleigh fading and the path-loss between the corresponding
transceivers. The coherence bandwidth of each wireless link is assumed
to be sufficiently high, so that each individual subcarrier block
experiences frequency flat fading, although the level of fading may
vary from one subcarrier block to another
in each transmission period. Additionally, the transceivers are stationary
or moving slowly enough so that the level of fading may
be assumed to be fixed for the duration of a scheduled
transmission period. Furthermore, the RAs are spaced sufficiently
far apart, so that each TA/RA pair experiences independent and identically
distributed~(i.i.d.) fading. Since these channels are slowly varying,
the system is capable of exploiting the benefits of channel
reciprocity associated with time-division duplexing~(TDD),
so that the CSI becomes available at each BS- and RN-transmitter and
at each possible RN- and UE-receiver. To elaborate, $\mathbf{H}_{n,k}^{BU}$
and $\mathbf{H}_{n,m}^{BR}$ are known at the BS, $\mathbf{H}_{n,m}^{BR}$
and $\mathbf{H}_{n,m,k}^{RU}$ are known at the RN $m$, while $\mathbf{H}_{n,k}^{BU}$
and $\mathbf{H}_{n,m,k}^{RU}$ are also known at UE $k$. Additionally,
through the use of dedicated low-rate error-free feedback channels,
$\mathbf{H}_{n,m,k}^{RU}$ is also assumed to be known at the BS so
that the BS may perform network-wide scheduling%
\footnote{In this paper, since our focus is on the resource allocation and the
associated SE/ESE optimization problems, the idealized simplifying
assumption of the availability of perfect CSI is employed. At the
current stage, accounting for erroneous CSI using, for example, robust
optimization~\cite{Kim2008} is beyond the scope of this paper and
may be addressed in our future work.%
}. These channel matrices are assumed to have full row rank, which
may be achieved with a high probability for typical DL wireless channel
matrices.

Furthermore, each receiver suffers from additive white Gaussian noise~(AWGN)
having a power spectral density of $N_{0}$. The maximum instantaneous
transmission power available to the BS and to each RN due to regulatory
and health-constraints is $P_{max}^{B}$ and $P_{max}^{R}$, respectively.
Since OFDMA modulation constitutes a linear operation, we focus our
attention on a single subcarrier block and as usual, we employ the
commonly-used equivalent baseband signal model%
\footnote{Since the specific signal model expressions of each link is dependent
on the transmission protocol to be designed, they are not presented
here but instead detailed in Section~\ref{sec:TxProt}.%
}.

\section{Transmission protocol design\label{sec:TxProt}}

The system can simultaneously use two transmission modes to convey
information to the UEs, namely the BS-to-UE mode, and the relaying-based
BS-to-RN and RN-to-UE mode. Note that although in classic OFDMA each
data stream is orthogonal in frequency, for the sake of further improving
the system's attainable SE or ESE performance, our system employs
spatial multiplexing in conjunction with ZFBF \emph{so that multiple
data streams may be served using the same subcarrier block, without
suffering from interference}. Additionally, since the relaying-based
transmission can be split into two phases, the design philosophy of
the BF matrices in each phase are described separately, although for
simplicity we have assumed that the respective channel matrices remain
unchanged in both phases. Firstly,
the definition of the semi-orthogonality criterion is given as follows~\cite{Yoo2006}. 
\begin{defn}
A pair of MISO channels, represented by the complex-valued column
vectors $\mathbf{v}_{1}$ and $\mathbf{v}_{2}$, are said to be semi-orthogonal
to each other with parameter $\alpha\in\left[0,1\right]$, when 
\begin{equation}
\frac{\left|\Re\left(\mathbf{v}_{1}^{\mathrm{H}}\mathbf{v}_{2}\right)\right|}{\|\mathbf{v}_{1}\|\|\mathbf{v}_{2}\|}\leq\alpha.\label{eq:semi-ortho}
\end{equation}
To be more specific, a measure of the grade of orthogonality between
$\mathbf{v}_{1}$ and $\mathbf{v}_{2}$ is given by the left-hand
side of inequality~(\ref{eq:semi-ortho}), which ranges from $0$
for orthogonal vectors to $1$ for linearly dependent vectors. 
\end{defn}
The authors of~\cite{Yoo2006} demonstrated that employing the ZFBF
strategy for MISO channels that satisfy $\alpha\to0$, while the number
of users obeys $K\to\infty$, asymptotically achieves the DPC capacity,
and it is therefore optimal for the BC channel. Similar principles
are followed when maximizing the SE or ESE of the
system considered in this paper.

\subsection{BF design for the first transmission phase\label{sub:BFT1}}

In the first transmission phase, only the BS is transmitting, while
both the RNs and the UEs act as receivers. This is similar to the
classic DL multi-user MIMO model. As described above, our aim is 1)
to design a ZFBF matrix for the BS to avoid interference between data
streams, and 2) to design receive BF matrices for
the UEs and RNs so that the resultant \emph{effective DL channel matrices}
contain as many semi-orthogonal
rows as possible that satisfy~(\ref{eq:semi-ortho})
for a given $\alpha$. Ideally, all receivers~(UEs and RNs) should
jointly compute%
\footnote{The joint computation is required only for attaining the highest number
of semi-orthogonal rows globally.%
} their receive BF matrices to accomplish the second
goal. However, this is generally impossible, since we cannot realistically
assume that the channel matrices associated with
each UE and RN are shared among them, due to the geographically-distributed
nature of the UEs and RNs. As a compromise, we opt for guaranteeing
that each individual effective DL channel
matrix contains locally orthogonal rows by employing
the SVD~\cite{Yoo2006,UlHassan2009}. Although these locally orthogonal rows may not remain orthogonal
globally, they can be characterized using the semi-orthogonality metric
of~(\ref{eq:semi-ortho}).

\begin{figure}
\begin{centering}
\includegraphics[scale=0.7]{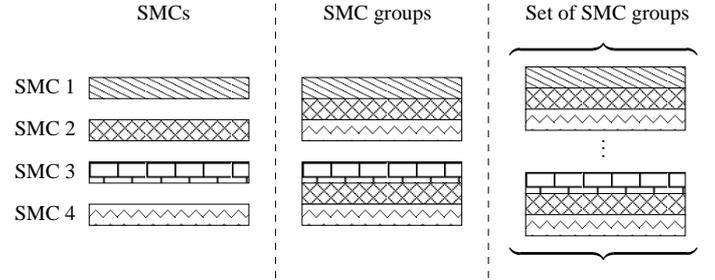} 
\par\end{centering}

\protect\caption{A conceptual illustration of the differences between SMCs, SMC groups
and a set of SMC groups.}

\label{fig:SMCs} 
\end{figure}

Bearing this in mind, the channel matrices of all DL transmissions
originating from the BS are decomposed at the BS,
UEs and RNs using the SVD~\cite{Boyd2004} as $\mathbf{H}_{n,k}^{BU}=\mathbf{U}_{n,k}^{BU}\mathbf{S}_{n,k}^{BU}\left(\mathbf{V}_{n,k}^{BU}\right)^{\mathrm{H}}$
and $\mathbf{H}_{n,m}^{BR}=\mathbf{U}_{n,m}^{BR}\mathbf{S}_{n,m}^{BR}\left(\mathbf{V}_{n,m}^{BR}\right)^{\mathrm{H}}$,
respectively. Thus, the receive-BF matrices for UE $k$ and RN $m$
are given by $\mathbf{R}_{n,k}^{BU,T_{1}}=\left(\mathbf{U}_{n,k}^{BU}\right)^{\mathrm{H}}\in\mathbb{C}^{N_{U}\times N_{U}}$
and $\mathbf{R}_{n,m}^{BR,T_{1}}=\left(\mathbf{U}_{n,m}^{BR}\right)^{\mathrm{H}}\in\mathbb{C}^{N_{R}\times N_{R}}$,
and the effective DL channel matrices are then given%
\footnote{Note that $T_{1}$ is used for indicating the first transmission phase,
and underline is used to denote the effective DL
channel matrices.%
} by $\underline{\mathbf{H}}_{n,k}^{BU,T_{1}}=\mathbf{R}_{n,k}^{BU,T_{1}}\mathbf{H}_{n,k}^{BU}=\mathbf{S}_{n,k}^{BU}\left(\mathbf{V}_{n,k}^{BU}\right)^{\mathrm{H}}\in\mathbb{C}^{N_{U}\times N_{B}}$
and $\underline{\mathbf{H}}_{n,m}^{BR,T_{1}}=\mathbf{R}_{n,m}^{BR,T_{1}}\mathbf{H}_{n,m}^{BR}=\mathbf{S}_{n,m}^{BR}\left(\mathbf{V}_{n,m}^{BR}\right)^{\mathrm{H}}\in\mathbb{C}^{N_{R}\times N_{B}}$,
respectively. Since $\mathbf{V}_{n,k}^{BU}$ and $\mathbf{V}_{n,m}^{BR}$
are both unitary, while $\mathbf{S}_{n,k}^{BU}$ and $\mathbf{S}_{n,m}^{BR}$
are both real and diagonal, these effective DL
channel matrices respectively consist of $\min\left(N_{B},N_{U}\right)$
and $\min\left(N_{B},N_{R}\right)$ orthogonal non-zero rows
\footnote{The reason why we use $\min(N_{B},N_{U})$ and $\min(N_{B},N_{R})$,
instead of $N_{U}$ and $N_{R}$, is because the antenna configuration
$N_{B}\le N_{U}$ and/or $N_{B}\le N_{R}$ is also covered.} with norms equal to their corresponding singular values. We refer
to these non-zero orthogonal rows as the SMCs of their associated
MIMO channel matrix
\footnote{Note that only when $N_{B}\ge N_{U}$ and $N_{B}\ge N_{R}$, a single
SMC is generated for each receive antenna.}. The $K$ BS-to-UE MIMO channel matrices and $M$ BS-to-RN channel
matrices generate a total of $\left[K\cdot\min\left(N_{B},N_{U}\right)+M\cdot\min\left(N_{B},N_{R}\right)\right]$
SMCs. Since these SMCs are generated from independent MIMO channel
matrices associated with geographically distributed UEs and RNs, they
are not all guaranteed to be orthogonal to each other. Furthermore,
since each UE or RN has multiple antennas and $N_{B}$ might not be
sufficiently large to simultaneously support all UEs and RNs, we have
to determine which specific SMCs should be served. As a result, for
each two-phase transmission period, we opt for selecting a SMC group
accounting for both phases from the set of available SMC groups. This
selection process is achieved by jointly using the SMC grouping algorithm
and solving the optimization problem detailed below. For the sake
of clarity, the concepts of the SMC, of the SMC group and of the set
of SMC groups are illustrated in Fig.~\ref{fig:SMCs}.

To elaborate a little further, a set of SMC groups, $\mathcal{G}_{n}$,
which is associated with subcarrier block $n$,
may be obtained using one of the grouping algorithms presented in
Section~\ref{sec:SGAlgor}. The BS selects a single group, $j\in\mathcal{G}_{n}$,
containing~(but not limited to%
\footnote{The SMC group selection, as a part of the scheduling operation, is
carried out at the BS before initiating the first transmission phase.
Hence, the selected SMC group will also contain $Q_{j}^{T_{2}}$ SMCs
selected by the BS for the second transmission phase, as detailed
in Section~\ref{sub:BFT2}.
} $Q_{j}^{T_{1}}$ SMCs out of the $\left[K\cdot\min\left(N_{B},N_{U}\right)+M\cdot\min\left(N_{B},N_{R}\right)\right]$
available SMCs to be supported by using ZFBF. Thus, we have $Q_{j}^{T_{1}}\leq\min\left[N_{B},K\cdot\min\left(N_{B},N_{U}\right)+M\cdot\min\left(N_{B},N_{R}\right)\right]$
and a multiplexing gain of $Q_{j}^{T_{1}}$ is achieved.
Let us denote the refined effective DL channel matrix
with rows being the $Q_{j}^{T_{1}}$ selected SMCs as $\underline{\mathbf{H}}_{n,j}^{T_{1}}\in\mathbb{C}^{Q_{j}^{T_{1}}\times N_{B}}$.
The ZFBF transmit matrix applied at the BS to subcarrier block $n$
is then given by the following right inverse $\mathbf{T}_{n,j}^{T_{1}}=\left(\underline{\mathbf{H}}_{n,j}^{T_{1}}\right)^{\mathrm{H}}\cdot\left[\underline{\mathbf{H}}_{n,j}^{T_{1}}\left(\underline{\mathbf{H}}_{n,j}^{T_{1}}\right)^{\mathrm{H}}\right]^{-1}\in\mathbb{C}^{N_{B}\times Q_{j}^{T_{1}}}$.
Since $\underline{\mathbf{H}}_{n,j}^{T_{1}}\mathbf{T}_{n,j}^{T_{1}}=\mathbf{I}_{N_{B}}$,
the potential interference between the $Q_{j}^{T_{1}}$
selected SMCs is completely avoided. Furthermore,
the columns of $\mathbf{T}_{n,j}^{T_{1}}$ are normalized by multiplying
the diagonal matrix $\mathbf{W}_{n,j}^{T_{1}}$ on the right-hand
side of $\mathbf{T}_{n,j}^{T_{1}}$ to ensure that each SMC transmission
is initially set to unit power%
\footnote{Each diagonal element of $\mathbf{W}_{n,j}^{T_{1}}$ is equal to the
reciprocal of the norm of the column vector to be normalized.%
}.

Then, $\mathbf{T}_{n,j}^{T_{1}}\mathbf{W}_{n,j}^{T_{1}}$ is used
as the DL transmit-BF matrix for the BS in the first
phase. Thus, the effective channel-to-noise ratios~(CNRs) in the
first transmission phase can be written as $G_{n,j,e_{1}}^{BU,T_{1}}=\left|w_{n,j,e_{1}}^{BU,T_{1}}\right|^{2}/\Delta\gamma N_{0}W$
and $G_{n,j,e}^{BR,T_{1}}=\left|w_{n,j,e}^{BR,T_{1}}\right|^{2}/\Delta\gamma N_{0}W$,
respectively, where $w_{n,j,e_{1}}^{BU,T_{1}}$ and $w_{n,j,e}^{BR,T_{1}}$
are the diagonal elements in $\mathbf{W}_{n,j}^{T_{1}}$. More specifically,
these diagonal elements correspond to SMC group $j$ and subcarrier
block $n$, and they are associated with either a direct BS-to-UE
SMC  or a BS-to-RN SMC. The additional subscripts
$e_{1}\in\{0,\cdots,\min[N_{B},K\cdot\min(N_{B},N_{U})]\}$ and $e\in\{0,\cdots,\min[N_{B},M\cdot\min(N_{B},N_{R}),K\cdot\min(N_{R},N_{U})]\}$
are used for distinguishing the multiple selected SMCs of the direct
links (i.e. those related to UEs), from the multiple selected SMC-pairs%
\footnote{A single SMC-pair consists of a SMC for the first phase and another
for the second phase. Although these SMCs are generated separately
in each phase, the SMC-pair associated with a common RN has to be
considered as a single entity in the SMC grouping algorithms presented
in Section \ref{sec:SGAlgor}. %
} that may be associated with a particular RN $\mathcal{M}(e)$, respectively.
Note that $\mathcal{M}\left(e\right)$ is a function of $e$, representing
the RN index~(similar to $m$ used before) associated with the SMC-pair
$e$, as further detailed in Section~\ref{sec:SGAlgor}.

At a given bit-error rate~(BER) requirement, $\Delta\gamma$ is the
signal-to-noise ratio~(SNR) gap between the lower-bound
SNR required for achieving the discrete-input--continuous-output memoryless
channel~(DCMC) capacity and the actual higher SNR
required by the modulation/coding schemes of the practical physical
layer transceivers employed. For example, making the simplifying assumption
that idealized transceivers capable of achieving
exactly the DCMC capacity are employed, then $\Delta\gamma=0\mbox{ dB}$.
Although, strictly speaking, so far it is not possible
to operate exactly at the DCMC channel capacity, there does exist several
physical layer transceiver designs that operate very close
to it~\cite{Hanzo2009}. Furthermore, the noise power received on
each subcarrier block is given by $N_{0}W$.

\subsection{BF design in the second transmission phase\label{sub:BFT2}}

The second transmission phase may be characterized
by the \textit{MIMO interference channel}. A similar methodology is
employed in the second transmission phase, except that now both the
BS and the RNs are transmitters, while a number of UEs are receiving.
In this phase, our aim is 1) to design ZFBF matrices for the BS and
RNs to avoid interference between data streams, 2) and to design
a receive-BF matrix for each UE so that the effective channel matrices
associated with each of its transmitters contain
rows which satisfy the semi-orthogonal condition~(\ref{eq:semi-ortho})
for a given $\alpha$. This means that more data streams may be served
simultaneously, thus improving the attainable SE or ESE performance.
Since there are multiple \textit{distributed} transmitters/MIMO
channel matrices associated with each UE, the SVD method described
in Section~\ref{sub:BFT1}, which is performed in a centralized fashion,
cannot be readily applied at the transmitter side.
Instead, we aim for minimizing the resultant correlation between the
generated SMCs, thus increasing the number of SMCs which satisfy~(\ref{eq:semi-ortho})
for a given $\alpha$. To accomplish this goal, we begin by introducing
the shorthand of $\underline{\mathbf{H}}_{n,k}^{BU,T_{2}}=\mathbf{R}_{n,k}^{U,T_{2}}\mathbf{H}_{n,k}^{BU}\in\mathbb{C}^{N_{U}\times N_{B}}$
and $\underline{\mathbf{H}}_{n,m,k}^{RU,T_{2}}=\mathbf{R}_{n,k}^{U,T_{2}}\mathbf{H}_{n,m,k}^{RU}\in\mathbb{C}^{N_{U}\times N_{R}}$
as the effective channel matrices between the BS and UE $k$, and
between RN $m$ and UE $k$, respectively, on subcarrier block $n$
in the second transmission phase, where $\mathbf{R}_{n,k}^{U,T_{2}}\in\mathbb{C}^{N_{U}\times N_{U}}$
is the yet-to-be-determined UE $k$'s receive-BF matrix. In light
of the preceding discussions, one of our aims is to design $\mathbf{R}_{n,k}^{U,T_{2}}$
so that the off-diagonal values of the matrices given by $\mathbf{A}_{0}=\underline{\mathbf{H}}_{n,k}^{BU,T_{2}}\left(\underline{\mathbf{H}}_{n,k}^{BU,T_{2}}\right)^{\mathrm{H}}=\mathbf{R}_{n,k}^{U,T_{2}}\mathbf{H}_{n,k}^{BU}\left(\mathbf{H}_{n,k}^{BU}\right)^{\mathrm{H}}\left(\mathbf{R}_{n,k}^{U,T_{2}}\right)^{\mathrm{H}}$
and $\mathbf{A}_{m}=\underline{\mathbf{H}}_{n,m,k}^{RU,T_{2}}\left(\underline{\mathbf{H}}_{n,m,k}^{RU,T_{2}}\right)^{\mathrm{H}}=\mathbf{R}_{n,k}^{U,T_{2}}\mathbf{H}_{n,m,k}^{RU}\left(\mathbf{H}_{n,m,k}^{RU}\right)^{\mathrm{H}}\left(\mathbf{R}_{n,k}^{U,T_{2}}\right)^{\mathrm{H}}\mbox{, }\forall m$
are as small as possible. This design goal may be formalized as 
\begin{eqnarray}
\underset{\mathbf{R}_{n,k}^{U,T_{2}}}{\mbox{min. }} &  & \left|\left|\mathbf{H}_{n,k}^{BU}\left(\mathbf{H}_{n,k}^{BU}\right)^{\mathrm{H}}-\left(\mathbf{R}_{n,k}^{U,T_{2}}\right)^{-1}\boldsymbol{\Lambda}_{0}\left(\mathbf{R}_{n,k}^{U,T_{2}}\right)^{\mathrm{-H}}\right|\right|_{\mathrm{F}}^{2}\nonumber \\
 &  & +\sum_{m=1}^{M}\left|\left|\mathbf{H}_{n,m,k}^{RU}\left(\mathbf{H}_{n,m,k}^{RU}\right)^{\mathrm{H}}\right.\right.\nonumber \\
 &  & \left.\left.-\left(\mathbf{R}_{n,k}^{U,T_{2}}\right)^{-1}\boldsymbol{\Lambda}_{m}\left(\mathbf{R}_{n,k}^{U,T_{2}}\right)^{\mathrm{-H}}\right|\right|_{\mathrm{F}}^{2},\label{eq:diagonalization}
\end{eqnarray}
where $\boldsymbol{\Lambda}_{0}$ and $\boldsymbol{\Lambda}_{m}$
are diagonal matrices containing the diagonal elements of $\mathbf{A}_{0}$
and $\mathbf{A}_{m}$, respectively. Therefore, $\left(\mathbf{R}_{n,k}^{U,T_{2}}\right)^{-1}$
is the \emph{jointly diagonalizing matrix}~\cite{Yeredor2002}, while
$\mathbf{H}_{n,k}^{BU}\left(\mathbf{H}_{n,k}^{BU}\right)^{\mathrm{H}}$
and $\mathbf{H}_{n,m,k}^{RU}\left(\mathbf{H}_{n,m,k}^{RU}\right)^{\mathrm{H}}$,
$\forall m$ are the matrices to be diagonalized. Thus, the algorithm
presented in~\cite{Yeredor2002} for solving%
\footnote{\label{footnote:joint_diag}In fact, when there are only two matrices
to diagonalize, say $\mathbf{A}_{0}$ and $\mathbf{A}_{1}$, the diagonalizing
matrix may be obtained from the eigenvectors of $\mathbf{A}_{0}\left(\mathbf{A}_{1}\right)^{-1}$~\cite{Comon2010}.
This diagonalizing matrix is able to fully diagonalize both $\mathbf{A}_{0}$
and $\mathbf{A}_{1}$.%
}~(\ref{eq:diagonalization}) may be invoked at
UE $k$ for obtaining $\mathbf{R}_{n,k}^{U,T_{2}}$, which may be
further fed back to the BS and RNs. Hence, the BS and RNs do not have
to share ${\bf H}_{n,k}^{BU}$ or ${\bf H}_{n,k}^{RU}$ via the wireless
channel and do not have to solve~(\ref{eq:diagonalization}) again.
As a result, we accomplish the goal of creating effective
channel matrices that contain rows aiming to satisfy~(\ref{eq:semi-ortho}).
Additionally, the columns of $\mathbf{R}_{n,k}^{U,T_{2}}$
have been normalized so that the power assigned for each SMC remains
unaffected.

After obtaining the receive-BF matrix, the SMCs of the transmissions
to UE $k$ on subcarrier block $n$ are given by
the non-zero rows of the effective channel matrices $\underline{\mathbf{H}}_{n,k}^{BU,T_{2}}$
and $\underline{\mathbf{H}}_{n,m,k}^{RU,T_{2}}$, $\forall m$. Since
the BS and the RNs act as distributed broadcasters in the second phase,
they are only capable of employing \textit{separate}
ZFBF transmit matrices to ensure that none of them imposes interference
on the SMCs it does not explicitly intend to serve. By employing one
of the grouping algorithms described in Section~\ref{sec:SGAlgor},
the BS schedules%
\footnote{\label{footnote:auxiliary_SMC}To elaborate a little further, when
computing its ZFBF transmit matrix, each transmitter~(either the BS
or a RN) must take into account an auxiliary SMC,
which is also selected from the legitimate SMC candidates and is required
for nulling the interference that this particular transmitter imposes
on each selected information-bearing SMC of the other transmitters.
Furthermore, each auxiliary SMC is employed by its corresponding transmitter
to transmit several additional zeros that are padded to the normal
data symbols. As a beneficial result, no interference is received
at each UE from the transmitter that does not serve this particular
UE. For more details of the SMC-based transmission in the second phase,
please refer to Algorithm~\ref{algo:SMC_check} described in Section~\ref{subsec:SMC_checking}.}
 $Q_{j}^{T_{2}}\leq\min\left[\min\left(N_{B},N_{R}\right),\sum_{i=1}^{K}L_{i}^{B}+L_{i}^{R}\right]$
SMCs to serve simultaneously in the second phase, where $L_{i}^{B}$
and $L_{i}^{R}$ represent the number of SMCs of UE
$i$ served by the BS and by RNs in this phase, respectively, where
we have $L_{i}^{B}+L_{i}^{R}\le N_{U}$, $L_{i}^{B}\le\min(N_{B},N_{U})$,
and $L_{i}^{R}\le\min(N_{R},N_{U})$. Note that since UE $i$ may
be simultaneously served both by the BS and by a RN~(each of them
serves a fraction of UE $i$'s SMCs), it is \textit{possible}
that the summation of the respective number of UEs served
\footnote{If at least one SMC of a UE is served by the BS~(or a RN), we say
that this UE is served by the BS~(or the RN). %
} by the BS and by RNs may be higher than $K$. Let
us denote the \textit{refined} effective DL channel matrices, from
the perspectives of the BS and RN $m$, consisting of the $Q_{j}^{T_{2}}$
selected SMCs as $\underline{\mathbf{H}}_{n,j}^{B,T_{2}}\in\mathbb{C}^{Q_{j}^{T_{2}}\times N_{B}}$
and $\underline{\mathbf{H}}_{n,j,m}^{R,T_{2}}\in\mathbb{C}^{Q_{j}^{T_{2}}\times N_{R}}$,
respectively. Since these are known to each transmitter, they may
employ ZFBF transmit matrices in the second phase, given by the right
inverses $\mathbf{T}_{n,j}^{B,T_{2}}=\left(\underline{\mathbf{H}}_{n,j}^{B,T_{2}}\right)^{\mathrm{H}}\cdot\left[\underline{\mathbf{H}}_{n,j}^{B,T_{2}}\left(\underline{\mathbf{H}}_{n,j}^{B,T_{2}}\right)^{\mathrm{H}}\right]^{-1}\in\mathbb{C}^{N_{B}\times Q_{j}^{T_{2}}}$
for the BS, and $\mathbf{T}_{n,j,m}^{R,T_{2}}=\left(\underline{\mathbf{H}}_{n,j,m}^{R,T_{2}}\right)^{\mathrm{H}}\cdot\left[\underline{\mathbf{H}}_{n,j,m}^{R,T_{2}}\left(\underline{\mathbf{H}}_{n,j,m}^{R,T_{2}}\right)^{\mathrm{H}}\right]^{-1}\in\mathbb{C}^{N_{R}\times Q_{j}^{T_{2}}}$
for RN $m$. Similar to the first transmission phase,
these ZFBF transmit matrices are normalized by $\mathbf{W}_{n,j}^{BU,T_{2}}$
and $\mathbf{W}_{n,j,m}^{RU,T_{2}}$, respectively, to ensure that
each SMC transmission is initially set to unit power. Upon
obtaining the selected SMCs, we denote the effective
CNRs in the second transmission phase as $G_{n,j,e_{2}}^{BU,T_{2}}=\left|w_{n,j,e_{2}}^{BU,T_{2}}\right|^{2}/\Delta\gamma N_{0}W$
and $G_{n,j,e}^{RU,T_{2}}=\left|w_{n,j,e}^{RU,T_{2}}\right|^{2}/\Delta\gamma N_{0}W$,
where $w_{n,j,e_{2}}^{BU,T_{2}}$ and $w_{n,j,e}^{RU,T_{2}}$ are
the diagonal elements in $\mathbf{W}_{n,j}^{BU,T_{2}}$
and $\mathbf{W}_{n,j,\mathcal{M}(e)}^{RU,T_{2}}$, respectively, and
the subscript $\mathcal{M}(e)$ has been defined in Section~\ref{sub:BFT1}.
To elaborate, for a second-phase BS-to-UE link, $w_{n,j,e_{2}}^{BU,T_{2}}$
corresponds to SMC group $j$ and subcarrier block $n$, while the
subscript $e_{2}\in\{0,\cdots,\min[N_{B},K\cdot\min(N_{B},N_{U})]\}$
is employed for further distinguishing the multiple selected SMCs
associated with UEs from the BS. Similarly, $w_{n,j,e}^{RU,T_{2}}$,
which also corresponds to SMC group $j$ and subcarrier
block $n$, is associated with the second-phase RN-to-UE link between
RN $\mathcal{M}\left(e\right)$ and the particular UE determined by
the SMC-pair $e$. 
\begin{figure}
\begin{centering}
\includegraphics[scale=0.57]{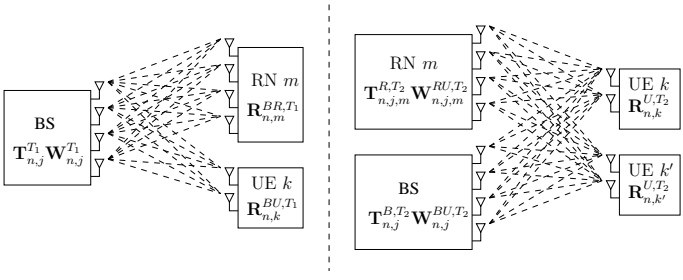} 
\par\end{centering}

\protect\caption{A conceptual schematic of the transmit- and receive-BF matrices employed
in the first and second transmission phases. In the first phase, the
BS applies the ZFBF transmit matrix $\mathbf{T}_{n,j}^{T_{1}}$ in
order to serve multiple data streams without imposing interference
between them. In the second phase, the BS and RNs employ separate
ZFBF matrices to distributively avoid interference between the data
streams being served.}

\label{fig:TS} 
\end{figure}

For more explicit clarity, a schematic of the transmit and receive
beamforming matrices in the first and second transmission phases is
presented in Fig.~\ref{fig:TS}.

\subsection{Achievable spectral efficiency and energy spectral efficiency\label{sub:AchieveSE}}

\begin{figure*}
\begin{equation}
C_{T}\left(\mathcal{P},\mathcal{S}\right)=\sum_{n=1}^{N}\sum_{j\in\mathcal{G}_{n}}s_{n,j}\left[\sum_{e_{1}\in\mathcal{E}_{n,j}}C_{n,j,e_{1}}^{BU,T_{1}}\left(\mathcal{P}\right)+\sum_{e_{2}\in\mathcal{E}_{n,j}}C_{n,j,e_{2}}^{BU,T_{2}}\left(\mathcal{P}\right)+\sum_{e\in\mathcal{E}_{n,j}}C_{n,j,e}^{BRU}\left(\mathcal{P}\right)\right].\label{eq:C_T}
\end{equation}
\begin{eqnarray}
P_{T}\left(\mathcal{P},\mathcal{S}\right) & = & \left(P_{C}^{B}+M\cdot P_{C}^{R}\right)+\frac{1}{2}\sum_{n=1}^{N}\sum_{j\in\mathcal{G}_{n}}s_{n,j}\Bigg[\xi^{B}\left(\sum_{e_{1}\in\mathcal{E}_{n,j}}P_{n,j,e_{1}}^{BU,T_{1}}+\sum_{e_{2}\in\mathcal{E}_{n,j}}P_{n,j,e_{2}}^{BU,T_{2}}\right)\nonumber \\
 &  & +\sum_{e\in\mathcal{E}_{n,j}}\left(\xi^{B}P_{n,j,e}^{BR,T_{1}}+\xi^{R}P_{n,j,e}^{RU,T_{2}}\right)\Bigg].\label{eq:P_T}
\end{eqnarray}
\hrulefill{} 
\end{figure*}

For the sake of convenience, let us first denote
the transmit power allocation policy as $\mathcal{P}$, which is a
set composed by all transmit power control variables invoked at the
BS and/or RNs in both transmission phases. Since
receive-BF is employed in conjunction with ZFBF, each SMC transmission
may be viewed as a single-input--single-output~(SISO) link. Therefore,
on the direct links, the receiver's SNR at UE $k$ corresponding to
SMCs $e_{1}$ and $e_{2}$ may be expressed as $\Gamma_{n,j,e_{1}}^{BU,T_{1}}\left(\mathcal{P}\right)=G_{n,j,e_{1}}^{BU,T_{1}}P_{n,j,e_{1}}^{BU,T_{1}}$
and $\Gamma_{n,j,e_{2}}^{BU,T_{2}}\left(\mathcal{P}\right)=G_{n,j,e_{2}}^{BU,T_{2}}P_{n,j,e_{2}}^{BU,T_{2}}$
for the first and second transmission phases, respectively. The
scalar variables $P_{n,j,e_{1}}^{BU,T_{1}}$ and $P_{n,j,e_{2}}^{BU,T_{2}}$,
which are elements of $\mathcal{P}$, determine the
transmit power values for SMCs $e_{1}$ and $e_{2}$ on the direct
links.  As a result, the achievable instantaneous
SE of the direct links is given by $C_{n,j,e_{1}}^{BU,T_{1}}\left(\mathcal{P}\right)=\frac{1}{2}\log_{2}\left(1+\Gamma_{n,j,e_{1}}^{BU,T_{1}}\left(\mathcal{P}\right)\right)$
and $C_{n,j,e_{2}}^{BU,T_{2}}\left(\mathcal{P}\right)=\frac{1}{2}\log_{2}\left(1+\Gamma_{n,j,e_{2}}^{BU,T_{2}}\left(\mathcal{P}\right)\right)$,
which are normalized both by time and by frequency to give
units of $\left[\mbox{bits/sec/Hz}\right]$. The factor of $\frac{1}{2}$
accounts for the fact that the transmission period is split into two
phases.

Similarly, for the SMC-pair $e$ of the DF relaying
links, the SNR at RN $\mathcal{M}(e)$ in the first
transmission phase is given by $\Gamma_{n,j,e}^{BR,T_{1}}\left(\mathcal{P}\right)=G_{n,j,e}^{BR,T_{1}}P_{n,j,e}^{BR,T_{1}}$
and the SNR at UE $k$ in the second transmission
phase is formulated as $\Gamma_{n,j,e}^{RU,T_{2}}\left(\mathcal{P}\right)=G_{n,j,e}^{RU,T_{2}}P_{n,j,e}^{RU,T_{2}}$.
Additionally, $P_{n,j,e}^{BR,T_{1}}$ and $P_{n,j,e}^{RU,T_{2}}$
are also elements of $\mathcal{P}$. Since the RNs
employ the DF protocol, the achievable SE is limited by the weaker
of the two RN-related links~\cite{Laneman2004}
and is given by $C_{n,j,e}^{BRU}\left(\mathcal{P}\right)=\min\left[\frac{1}{2}\log_{2}\left(1+\Gamma_{n,j,e}^{BR,T_{1}}\left(\mathcal{P}\right)\right),\frac{1}{2}\log_{2}\left(1+\Gamma_{n,j,e}^{RU,T_{2}}\left(\mathcal{P}\right)\right)\right]$.

Let us now introduce the SMC group selection variable $s_{n,j}\in\left\{ 0,1\right\} $,
which indicates that SMC group $j$, as introduced in Sections~\ref{sub:BFT1}
and~\ref{sub:BFT2}, is selected for subcarrier block $n$, when
$s_{n,j}=1$, and $s_{n,j}=0$ otherwise. All SMC group selection
variables are scalars and are collected into a set denoted by $\mathcal{S}$.
Once again, we emphasize that $\mathcal{G}_{n}$
denotes the set of possible SMC groups for subcarrier block $n$.
Thus, the total achieved SE is given by~(\ref{eq:C_T}), where $\mathcal{E}_{n,j}$
is the set comprising the SMCs in the selected group $j$ on subcarrier
block $n$.

In this work, we adopt the energy dissipation model presented in~\cite{Auer2010},
where the total energy dissipation of the system is assumed to be
dependent on several factors, including the number of TAs, the energy
dissipation of the RF and baseband circuits, and the efficiencies
of the power amplifier, feeder cables, cooling system, mains power
supply, and converters. For the sake of simplicity, the total energy
dissipation as presented in~\cite{Auer2010} has been partitioned
into a fixed term, and a term that varies with the transmission powers.
Thus, the energy dissipation of the system may be characterized by~(\ref{eq:P_T}),
where $P_{C}^{B}$ and $P_{C}^{R}$ represent the fixed energy dissipation
of each BS and each RN, respectively, while $\xi^{B}>1$ and $\xi^{R}>1$
are the energy dissipation multipliers of the transmit
powers for the BS and the RNs, respectively. The effect of multiple
transmit antennas on the total energy dissipation has been included
in the terms $P_{C}^{B}$, $P_{C}^{R}$, $\xi^{B}$ and $\xi^{R}$.

Finally, the ESE of the system is expressed as 
\begin{equation}
\eta_{E}\left(\mathcal{P},\mathcal{S}\right)=\frac{C_{T}\left(\mathcal{P},\mathcal{S}\right)}{P_{T}\left(\mathcal{P},\mathcal{S}\right)}.\label{eq:eeff}
\end{equation}
The objective of this paper is to maximize~(\ref{eq:eeff}) by appropriately
optimizing $\mathcal{P}$ and $\mathcal{S}$.

\section{Semi-orthogonal grouping algorithms\label{sec:SGAlgor}}

As described in Section~\ref{sec:SysModel}, the BS has to choose
$Q_{j}^{T_{1}}$ and $Q_{j}^{T_{2}}$ SMCs for the first and second
transmission phases, respectively.
These selected SMCs collectively form the SMC group $j$.
Since the system supports both direct and relaying links, the grouping
algorithms described in~\cite{Yoo2006,UlHassan2009}, which were
designed for MIMO systems dispensing with relays, may not be directly
applied. Instead, we propose a pair of viable grouping algorithms,
namely the exhaustive search-based grouping algorithm~(ESGA), and
the orthogonal component-based grouping algorithm~(OCGA).

Note that because there are multiple distributed transmitters in the
second transmission phase, each UE designs its receive-BF matrix by
jointly considering all the MIMO channel matrices associated with
it, as described in Section~\ref{sub:BFT2}. However, before
applying this method, we have to determine which particular transmitters~(out
of the BS and RNs) should actively transmit in the second transmission
phase based on the results of SMC selection. Note that it is possible
that the SMC candidates obtained may lead to higher effective CNRs
when a subset of the transmitters are inactive. On the one hand, an
additional effect of only activating a subset of transmitters is the
reduced number of SMC candidates, which might in turn result in a
reduced number of qualified SMCs that satisfy the semi-orthogonality
criterion considered. As a result, the achievable spatial multiplexing
gain and SE might be degraded. On the other hand, this SE-reduction
effect may be counteracted by the improved CNRs gleaned from the fact
that it is easier to generate SMCs that can satisfy a stricter semi-orthogonality
criterion, specified by a smaller value of $\alpha$, when the number
of transmitters is lower. For example, in the scenarios where only
one or two active transmitters are selected, the UEs can employ receive-BF
matrices that create effective DL channel matrices containing completely
orthogonal rows by using the SVD or the exact diagonalization method~(see
Footnote \ref{footnote:joint_diag}), respectively. In order to account
for this dilemma, for the second transmission phase, the proposed
grouping algorithms evaluate a full list of SMCs,
which consists of the SMCs obtained from the $2^{M+1}-1$ possible
combinations of active transmitters~(the BS and $M$ RNs, while ignoring
the case when there are no active transmitters). Compared
to using a smaller list of SMCs, using a full list of SMCs ensures
that achieving a lower-bound SE is always guaranteed, while a higher
SE can only be obtained upon increasing the number of transmitters
in the system.

\subsection{SMC checking algorithm\label{subsec:SMC_checking}}

\begin{algorithm} {\scriptsize} \SetKw{Or}{or} \SetKw{True}{true} \SetKw{False}{false} \SetKwFunction{SMCCheck}{SMCCheck} \SetKwInOut{Input}{inputs}\SetKwInOut{Output}{outputs}
\Input{candidate SMC $e_{c}$, current SMC group $\mathcal{E}_{n,j}$, \newline semi-orthogonality parameter $\alpha$} \Output{\True or \False}
\BlankLine bool \SMCCheck$\left(e_{c},\mathcal{E}_{n,j},\alpha\right)$ \BlankLine \Begin{ 	\BlankLine 	$\mathcal{T}^{BS,T_{1}}\leftarrow \{\}$\; 	$\mathcal{T}^{BS,T_{2}}\leftarrow \{\}$\; 	$\mathcal{T}^{RN,T_{2}}_{m}\leftarrow \{\} \text{, } \forall m\in \{1,\cdots,M \}$\; 	$\mathcal{R}^{UE,T_{2}}_{k}\leftarrow \{\} \text{, } \forall k\in \{1,\cdots,K \}$\; 	\ForEach{SMC $e_{1}\in\mathcal{E}_{n,j}$} 	{\label{algo:SMC_check:start_trancount} 		$\mathcal{T}^{BS,T_{1}}\leftarrow\mathcal{T}^{BS,T_{1}}\cup\{e_{1}\}$\; 	} 	\ForEach{SMC $e_{2}\in\mathcal{E}_{n,j}$} 	{  		$\mathcal{T}^{BS,T_{2}}\leftarrow\mathcal{T}^{BS,T_{2}}\cup\{e_{2}\}$\; 		$\mathcal{R}_{k}^{UE,T_{2}}\leftarrow\mathcal{R}_{k}^{UE,T_{2}}\cup\{e_{2}\}$\;\label{algo:SMC_check:rececount} 	} 	\ForEach{SMC $e\in\mathcal{E}_{n,j}$}  	{  		$\mathcal{T}^{BS,T_{2}}\leftarrow\mathcal{T}^{BS,T_{2}}\cup\{e\}$\; 		$\mathcal{T}_{\mathcal{M}(e)}^{RN,T_{2}}\leftarrow\mathcal{T}_{\mathcal{M}(e)}^{RN,T_{2}}\cup\{e\}$\; 	}\label{algo:SMC_check:end_trancount} 
	\BlankLine 	\uIf{$e_{c}$ is BS transmission in $T_{1}$} 	{\label{algo:SMC_check:start_candidate} 		$\mathcal{T}^{BS,T_{1}}\leftarrow \mathcal{T}^{BS,T_{1}} \cup \{e_{c}\}$\; 		\lIf{$\mathcal{T}^{BS,T_{1}}$ is not $\alpha$-semi-orthogonal \Or $| \mathcal{T}^{BS,T_{1}}|>N_{B}$}{\label{algo:SMC_check:ortho_check1}\KwRet \False} 	} 	\uElseIf{$e_{c}$ is BS transmission in $T_{2}$} 	{ 		$\mathcal{T}^{BS,T_{2}}\leftarrow \mathcal{T}^{BS,T_{2}} \cup \{e_{c}\}$\; 		$\mathcal{T}^{RN,T_{2}}_{m}\leftarrow \mathcal{T}^{RN,T_{2}}_{m} \cup \{e^{\varnothing}_{m}\} \text{, } \forall m\in \{1,\cdots,M \}$\;\label{algo:SMC_check:aux1} 	} 	\ElseIf{$e_{c}$ is RN $m$ transmission in $T_{2}$} 	{ 		$\mathcal{T}^{RN,T_{2}}_{m}\leftarrow \mathcal{T}^{RN,T_{2}}_{m} \cup \{e_{c}\}$\; 		$\mathcal{T}^{BS,T_{2}}\leftarrow \mathcal{T}^{BS,T_{2}} \cup \{e^{\varnothing}\}$\;\label{algo:SMC_check:aux2} 		$\mathcal{T}^{RN,T_{2}}_{m'}\leftarrow \mathcal{T}^{RN,T_{2}}_{m'} \cup \{e^{\varnothing}_{m'}\} \text{, } \forall m'\in \{1,\cdots,M \} \setminus m$\;\label{algo:SMC_check:aux3} 	}\label{algo:SMC_check:end_candidate} 
	\BlankLine 	\lIf{$\mathcal{T}^{BS,T_{2}}$ is not $\alpha$-semi-orthogonal \Or $| \mathcal{T}^{BS,T_{2}}|>N_{B}$} 	{\label{algo:SMC_check:ortho_check2} 		\KwRet \False 	} 	\lIf{$\mathcal{T}^{RN,T_{2}}_{m}$ is not $\alpha$-semi-orthogonal \Or $| \mathcal{T}^{RN,T_{2}}_{m}|>N_{R} \text{, } m\in \{1,\cdots,M \}$} 	{\label{algo:SMC_check:ortho_check3} 		\KwRet \False 	} 
	\BlankLine 	\If{$e_{c}$ is UE $k$ reception in $T_{2}$} 	{ 		$\mathcal{R}^{UE,T_{2}}_{k}\leftarrow \mathcal{R}^{UE,T_{2}}_{k} \cup \{e_{c}\}$\;\label{algo:SMC_check:candidate_check} 		\lIf{$| \mathcal{R}^{UE,T_{2}}_{k}|>N_{U} \text{, } k \in \{1,\cdots,K\}$}{\label{algo:SMC_check:UE_count} \KwRet \False} 	} 
	\BlankLine 	\lIf{$|\mathcal{T}^{BS,T_{1}}|>\min\left(N_{B},KN_{U}+MN_{R}\right)$}{\label{algo:SMC_check:SMG_check1}\KwRet \False} 	\lIf{$\sum_{k=1}^{K}|\mathcal{R}^{UE,T_{2}}_{k}|>\min\left(N_{B},N_{R}\right)$}{\label{algo:SMC_check:SMG_check2}\KwRet \False} 
	\BlankLine 	\KwRet \True\; \label{algo:SMC_check:end_return} }
\caption{SMC checking algorithm} \label{algo:SMC_check} \end{algorithm}Both grouping algorithms must evaluate a particular SMC before it
may be included into the SMC group to be generated.
This evaluating and SMC-group updating process is depicted in Algorithm~\ref{algo:SMC_check}.
More specifically, the algorithm identifies the transmitters
associated with each SMC of the current SMC group, denoted by $\mathcal{E}_{n,j}$,
in lines~\ref{algo:SMC_check:start_trancount} to~\ref{algo:SMC_check:end_trancount}.
The transmitter associated with the candidate SMC, $e_{c}$, is identified
in lines~\ref{algo:SMC_check:start_candidate} to~\ref{algo:SMC_check:end_candidate}.
Additionally, as briefly pointed out in Footnote~\ref{footnote:auxiliary_SMC},
for an active transmitter, if the candidate SMC is associated with
a transmission in the second phase, then the auxiliary
SMCs, $e^{\varnothing}$ and $e_{m}^{\varnothing}$, are included
for the other active transmitters in lines~\ref{algo:SMC_check:aux1},~\ref{algo:SMC_check:aux2}
and~\ref{algo:SMC_check:aux3}, in order to ensure that these potentially
interfering transmitters do not impose interference on the candidate
SMC%
\footnote{For distributed transmitters encountered in the second transmission
phase, it is not feasible to design a single ZFBF transmit matrix
as we did for the BS in the first transmission phase. For the second
transmission phase, when $N_{B}\le N_{U}$ and $N_{R}\le N_{U}$,
each SMC is associated with a single receive antenna. Consider this
case as an example, when the BS is transmitting on a SMC to a particular
receive antenna of a UE, an active RN may be transmitting zeros on
an auxiliary SMC, which is also selected from the legitimate SMC candidates,
to the same receive antenna of that UE. As a beneficial result of
this strategy, for each transmitter, the interference imposed by other
active transmitters are nulled.%
}. Note that $e^{\varnothing}$ and $e_{m}^{\varnothing}$ represent
auxiliary SMCs invoked by the BS and RNs, respectively. Having
determined the transmitters associated with the SMCs, the algorithm
checks that the SMCs associated with the same transmitter satisfy
the semi-orthogonality criterion of~(\ref{eq:semi-ortho}) having
parameter $\alpha$ in lines~\ref{algo:SMC_check:ortho_check1},~\ref{algo:SMC_check:ortho_check2}
and~\ref{algo:SMC_check:ortho_check3}. Furthermore, the algorithm
ensures that the inclusion of the
candidate SMC does not force any of the transmitters to transmit over
its maximum number of transmit dimensions, as depicted in lines~\ref{algo:SMC_check:ortho_check2}
and~\ref{algo:SMC_check:ortho_check3}. Meanwhile,
each UE should not receive more than its maximum number of receive
dimensions, which is accomplished in lines~\ref{algo:SMC_check:rececount},~\ref{algo:SMC_check:candidate_check}
and~\ref{algo:SMC_check:UE_count}. Finally, the maximum achievable
spatial multiplexing gain should not be exceeded in either the first
or second phase, which is ensured by lines~\ref{algo:SMC_check:SMG_check1}
and~\ref{algo:SMC_check:SMG_check2}. If all of these checks are
successful, the algorithm exits with a true condition in line~\ref{algo:SMC_check:end_return}.

\subsection{ESGA and OCGA}

\begin{algorithm} \SetKw{Or}{or} \SetKw{True}{true} \SetKw{False}{false} \SetKwFunction{SMCCheck}{SMCCheck} \SetKwFunction{ESGA}{ESGA} \SetKwInOut{Input}{inputs}\SetKwInOut{Output}{outputs}
\Input{set of SMC groups associated with subcarrier block $n$ (initialized as empty set), $\mathcal{G}_{n}$ \newline current SMC group (initialized as empty set), $\mathcal{E}_{n,j}$ \newline SMCs associated with subcarrier block $n$, $\mathcal{E}_{n}$ \newline semi-orthogonality parameter $\alpha$} \Output{none} \BlankLine void \ESGA$\left(\mathcal{G}_{n},\mathcal{E}_{n,j},\mathcal{E}_{n},\alpha\right)$ \BlankLine \Begin{ 	\BlankLine 	\ForEach{$e_{c}\in \mathcal{E}_{n} $} 	{ \label{algo:ESGA:enumerate} 		\If{\SMCCheck$\left(e_{c},\mathcal{E}_{n,j},\alpha\right)$ } 		{ \label{algo:ESGA:SMC_check} 			$\mathcal{E}'_{n,j'}\leftarrow \mathcal{E}_{n,j} \cup \{e_{c}\}$\; \label{algo:ESGA:new_group} 			$\mathcal{G}_{n} \leftarrow \mathcal{G}_{n} \cup \{\mathcal{E'}_{n,j'}\}$\;\label{algo:ESGA:append} 			\ESGA$\left(\mathcal{G}_{n},\mathcal{E}'_{n,j'},\mathcal{E}_{n}\setminus e_c,\alpha\right)$\; \label{algo:ESGA:recursion} 		} 	}\label{algo:end_ESGA_enumerate} 	\KwRet\; }
\caption{Exhaustive search-based grouping algorithm~(ESGA)} \label{algo:ESGA} \end{algorithm}\begin{algorithm} \SetKw{Or}{or} \SetKw{True}{true} \SetKw{False}{false} \SetKwFunction{SMCCheck}{SMCCheck} \SetKwFunction{OCGA}{OCGA} \SetKwFunction{NOC}{NOC} \SetKwInOut{Input}{inputs}\SetKwInOut{Output}{outputs}
\Input{set of SMC groups associated with subcarrier block $n$ (initialized as empty set), $\mathcal{G}_{n}$ \newline current SMC group (initialized as empty set), $\mathcal{E}_{n,j}$ \newline SMCs associated with subcarrier block $n$, $\mathcal{E}_{n}$ \newline semi-orthogonality parameter $\alpha$} \Output{none} \BlankLine void \OCGA$\left(\mathcal{G}_{n},\mathcal{E}_{n,j},\mathcal{E}_{n},\alpha\right)$ \BlankLine \Begin{ 	\BlankLine 	complete $\leftarrow$ \True\; 	$\mathcal{E}_{c}\leftarrow \{\}$\; \label{algo:OCGA:candidate_set} 	\BlankLine 	\ForEach{$e_{c}\in \mathcal{E}_{n} $} 	{ \label{algo:OCGA:enumerate} 		\If{\SMCCheck$\left(e_{c},\mathcal{E}_{n,j},\alpha\right)$ } 		{ \label{algo:OCGA:SMC_check} 			\uIf{$\left|\mathcal{E}_{n,j}\right|==0$} 			{\label{algo:empty_check} 				$\mathcal{E}'_{n,j'}\leftarrow \mathcal{E}_{n,j} \cup \{e_{c}\}$\; \label{algo:OCGA:new_set_begin} 				\OCGA$\left(\mathcal{G}_{n},\mathcal{E}'_{n,j'},\mathcal{E}_{n}\setminus e_c,\alpha\right)$\; 				\KwRet\; \label{algo:OCGA:new_set_end} 			} 			\Else 			{ 				$\mathcal{E}_{c}\leftarrow \mathcal{E}_{c} \cup \{e_{c}\}$\; \label{algo:OCGA:add_candidate} 				complete $\leftarrow$ \False\; 			} 		} 	}\label{algo:end_of_creatiing_candidate_group}
	\BlankLine 	\uIf{complete} 	{ 		$\mathcal{G}_{n} \leftarrow \{\mathcal{E}_{n,j}\}$\;\label{algo:OCGA:cannot_find}  	} 	\Else 	{ 		$\mathcal{E}'_{n,j'}\leftarrow \mathcal{E}_{n,j} \cup \underset{e_{c}\in\mathcal{E}_{c}}{\arg\max}$ \NOC$\left(e_{c},\mathcal{E}_{n,j}\right)$\; \label{algo:OCGA:best_candidate} 		\OCGA$\left(\mathcal{G}_{n},\mathcal{E}'_{n,j'},\mathcal{E}_{n}\setminus e_c,\alpha\right)$\; \label{algo:OCGA:recursive} 	}
\BlankLine \KwRet\; }
\caption{Orthogonal component-based grouping algorithm~(OCGA)} \label{algo:OCGA} \end{algorithm} We present our first grouping method in Algorithm~\ref{algo:ESGA}.
Simply put, the ESGA recursively creates new SMC groups by exhaustively
searching through all the possible combinations of SMCs and including
those that pass the SMC checking algorithm. To elaborate, in
the loop ranging from line~\ref{algo:ESGA:enumerate} to line~\ref{algo:end_ESGA_enumerate},
the algorithm searches through all the possible SMCs associated with
subcarrier block $n$, which are collectively denoted
by $\mathcal{E}_{n}$ and satisfy $e_{c}\in\mathcal{E}_{n}$. The
specific SMCs that satisfy the checks performed in
line~\ref{algo:ESGA:SMC_check} are appended to the current SMC group
in line~\ref{algo:ESGA:new_group}, and the resultant updated SMC
group $\mathcal{E}'_{n,j'}$  is appended to the
set of SMC groups obtained for subcarrier block $n$ in line~\ref{algo:ESGA:append}.
Additionally, $\mathcal{E}'_{n,j'}$ is used recursively in line~\ref{algo:ESGA:recursion}
for filling this group and for forming new groups. The computational
complexity of ESGA is dependent on the number of SMCs which are semi-orthogonal
to each other. The worst-case complexity is obtained when every SMC
satisfies the checks performed in line~\ref{algo:ESGA:SMC_check},
leading to a time-complexity (in terms of the number
of SMC groups generated) upper-bounded (not necessarily tight) by
$\mathcal{O}\left(\sum_{n=1}^{N}\left|\mathcal{E}_{n}\right|^{\theta}\right)$,
where
\begin{eqnarray}
\theta & = & \min\left[N_{B},K\cdot\min\left(N_{B},N_{U}\right)+M\cdot\min\left(N_{B},N_{R}\right)\right]\nonumber \\
 &  & +\min\left[\min\left(N_{B},N_{R}\right),\sum_{i=1}^{K}L_{i}^{B}+L_{i}^{R}\right].
\end{eqnarray}
 In other words, each subcarrier block may be treated independently.
For each subcarrier block, $\left|\mathcal{E}_{n}\right|$ SMCs must
be checked until the maximum multiplexing gain in both the first and
second phases has been attained.

The second algorithm, OCGA, is presented in Algorithm~\ref{algo:OCGA},
which aims to be a lower complexity alternative to ESGA. The OCGA
commences by creating a SMC candidate set $\mathcal{E}_{c}$,
whose elements satisfy the checks performed in Algorithm~\ref{algo:SMC_check},
in lines~\ref{algo:OCGA:candidate_set} to~\ref{algo:end_of_creatiing_candidate_group}.
More specifically, if the current SMC group $\mathcal{E}_{n,j}$
is empty, the algorithm can simply create a new SMC group containing
only the candidate SMC that has passed the SMC checks
of Algorithm~\ref{algo:SMC_check} in lines~\ref{algo:empty_check}
to~\ref{algo:OCGA:new_set_end}. If the SMC group is not empty, the
algorithm adds to it the particular SMC candidate that results in
the highest norm of the orthogonal component~(NOC), via the Gram-Schmidt
procedure~\cite{Yoo2006,UlHassan2009}, in line~\ref{algo:OCGA:best_candidate}.
This process is repeated until the maximum multiplexing gain in both
the first and second phases has been attained. When comparing the
NOCs obtained for the relaying links, the minimum of the NOCs obtained
from the BS-to-RN and RN-to-UE SMCs is used. This is because the information
conveyed on the relaying link is limited by the weaker of the two
transmissions, which is reflected in the effective channel gains quantified
by these norms. If no SMCs satisfy the checks of
line~\ref{algo:OCGA:SMC_check}, the current SMC group is complete,
and it is appended to the current set of SMC groups in line~\ref{algo:OCGA:cannot_find}.
Since new groups are only created when the current SMC group is empty,
this algorithm results in much fewer groups than ESGA. The algorithmic
time-complexity is given by $\mathcal{O}\left(\sum_{n=1}^{N}\left|\mathcal{E}_{n}\right|\right)$
as a single group is created for each initially-selected SMC.

Both grouping algorithms may be initialized with
an empty SMC group, $\mathcal{E}_{n,j}\leftarrow\left\{ \right\} $,
and an empty set of SMC groups, $\mathcal{G}_{n}\leftarrow\left\{ \right\} $
, so that they recursively create and fill SMC groups according to
their criteria. Additionally, a final step is performed
to remove the specific groups, which result in effective channel gains
that are less than or equal to that of another group, while having
the same transmitters. Therefore, this final step does not reduce
the attainable SE or ESE, but reduces the number of possible groups,
thus alleviating the computational complexity imposed by the optimization
algorithms of Section~\ref{sub:DualDecomp}.

\section{SEM/ESEM Problem Formulation and Solution\label{sec:ProbForm}}

Having obtained the set of SMC groups $\mathcal{G}_{n}$
for each subcarrier block $n$, in this section our aim is to find
the optimum power variables contained in $\mathcal{P}$ and optimum
SMC-group selection variables contained in $\mathcal{S}$,
so that~(\ref{eq:eeff}) is maximized. We commence by formulating
the problem of maximizing the SE of the system as~(\ref{eq:obj_orig})--(\ref{eq:chipsi_orig}).
\begin{figure*}
\begin{eqnarray}
\underset{\mathcal{P},\mathcal{S}}{\mbox{maximize}} &  & C_{T}\left(\mathcal{P},\mathcal{S}\right)\label{eq:obj_orig}\\
\mbox{subject to} &  & \sum_{j\in\mathcal{G}_{n}}s_{n,j}\leq1\mbox{, }\forall n,\label{eq:theta_orig}\\
 &  & \sum_{i=1}^{N}\sum_{j\in\mathcal{G}_{n}}s_{n,j}\left[\sum_{e_{1}\in\mathcal{E}_{n,j}}P_{n,j,e_{1}}^{BU,T_{1}}+\sum_{e\in\mathcal{E}_{n,j}}P_{n,j,e}^{BR,T_{1}}\right]\leq P_{max}^{B},\label{eq:zeta_orig}\\
 &  & \sum_{i=1}^{N}\sum_{j\in\mathcal{G}_{n}}s_{n,j}\sum_{e_{2}\in\mathcal{E}_{n,j}}P_{n,j,e_{2}}^{BU,T_{2}}\leq P_{max}^{B},\label{eq:omega_orig}\\
 &  & \sum_{i=1}^{N}\sum_{j\in\mathcal{G}_{n}}s_{n,j}\sum_{\begin{subarray}{c}
e\in\mathcal{E}_{n,j}\\
\mathcal{M}\left(e\right)=m
\end{subarray}}P_{n,j,e}^{RU,T_{2}}\leq P_{max}^{R}\mbox{, }\forall m,\label{eq:nu_orig}\\
 &  & s_{n,j}\in\left\{ 0,1\right\} \mbox{, }\forall n,j,\label{eq:sigmatau_orig}\\
 &  & P_{n,j,e_{1}}^{BU,T_{1}},P_{n,j,e}^{BR,T_{1}},P_{n,j,e_{2}}^{BU,T_{2}},P_{n,j,e}^{RU,T_{2}}\geq0\mbox{, }\forall n,j,e_{1},e_{2},e.\label{eq:chipsi_orig}
\end{eqnarray}
\hrulefill{} 
\end{figure*}

To elaborate,~(\ref{eq:obj_orig}) represents the
sum SE of the system, which is formulated in more detail as~(\ref{eq:C_T}).
The constraints~(\ref{eq:zeta_orig})--(\ref{eq:nu_orig}) ensure
that the maximum instantaneous transmission power constraint is never
exceeded in either of the two transmission phases
for the BS and the RNs, while the constraints~(\ref{eq:theta_orig})
and~(\ref{eq:sigmatau_orig}) ensure that only a single SMC
group is selected for each subcarrier block. Finally,~(\ref{eq:chipsi_orig})
restricts the power variables to be non-negative.

\subsection{Relaxed SEM problem}

Although the constraint~(\ref{eq:chipsi_orig}) is affine~(hence
convex) in the optimization variables contained in $\mathcal{P}$,
~(\ref{eq:theta_orig})--(\ref{eq:nu_orig}) are non-convex~\cite{Boyd2004},
because~(\ref{eq:sigmatau_orig}) imposes a binary constraint on
the problem. Furthermore, the objective function given by~(\ref{eq:obj_orig})
is not concave, since it is dependent
on the binary variables given by $\mathcal{S}$. Thus,~(\ref{eq:obj_orig})--(\ref{eq:chipsi_orig})
may be classified as a \emph{mixed-integer nonlinear programming}~(MINLP)\emph{
}problem, which may be solved using branch-and-bound methods~\cite{Bertsekas1999}.
However, these methods typically incur a computational complexity
that increases exponentially in
the number of discrete variables, which is undesirable for practical
implementations. To circumvent this initial setback, we introduce
the following auxiliary variables 
\begin{eqnarray}
\widetilde{P}_{n,j,e_{1}}^{BU,T_{1}} & = & P_{n,j,e_{1}}^{BU,T_{1}}\widetilde{s}_{n,j}\mbox{, }\forall n,j,e_{1},\label{eq:aux_p1}\\
\widetilde{P}_{n,j,e}^{BR,T_{1}} & = & P_{n,j,e}^{BR,T_{1}}\widetilde{s}_{n,j}\mbox{, }\forall n,j,e,\\
\widetilde{P}_{n,j,e_{2}}^{BU,T_{2}} & = & P_{n,j,e_{2}}^{BU,T_{2}}\widetilde{s}_{n,j}\mbox{, }\forall n,j,e_{2},\\
\widetilde{P}_{n,j,e}^{RU,T_{2}} & = & P_{n,j,e}^{RU,T_{2}}\widetilde{s}_{n,j}\mbox{, }\forall n,j,e,\label{eq:aux_p4}
\end{eqnarray}
\begin{equation}
\widetilde{C}_{n,j,e_{1}}^{BU,T_{1}}\mbox{, }\widetilde{C}_{n,j,e_{2}}^{BU,T_{2}}\mbox{ and }\widetilde{C}_{n,j,e}^{BRU}\mbox{, }\forall n,j,e_{1},e_{2},e,\label{eq:aux_rates}
\end{equation}
where we have relaxed%
\footnote{In~\cite{Yu2006}, such a relaxation results in a time-sharing solution
regarding each subcarrier. In this work, this relaxation
may be viewed as time-sharing of each subcarrier block, as multiple
SMC groups can then occupy a fraction of each subcarrier block in
time. Naturally, the relaxation means
that we do not accurately solve the original problem of~(\ref{eq:obj_orig})--(\ref{eq:chipsi_orig}).
However, as shown in~\cite{Ng2012c,Ng2012,Cheung2013}, the solution
to the original problem is still obtained with high
probability when using the dual decomposition method on the relaxed
problem~(as in this work) as the number of subcarriers tends to infinity.
It was shown that $8$ subcarriers is sufficient for
this to be true in the context of~\cite{Seong2006}, while we have
shown that $2$ subcarriers is sufficient in the context of~\cite{Cheung2013}.
} the binary constraint of~(\ref{eq:sigmatau_orig}) to give 
\begin{equation}
0\leq\widetilde{s}_{n,j}\leq1\mbox{, }\forall n,j,
\end{equation}
so that we may write~(\ref{eq:obj_orig})--(\ref{eq:chipsi_orig})
in the hypograph problem~\cite{Boyd2004} form given by~(\ref{eq:obj_relax})--(\ref{eq:chipsi_relax})%
\footnote{Writing the original optimization problem in the
hypograph form of~(\ref{eq:obj_relax})--(\ref{eq:chipsi_relax})
means that minimum per-link or system-wide SE constraints
may be readily introduced. However, minimum SE constraints are not
considered in this paper as our goal is to find the maximum SE/ESE
solutions, which may not be equivalent to the solutions
obtained when satisfying minimum SE constraints.
}, where $\widetilde{\mathcal{C}}$, $\widetilde{\mathcal{P}}$ and
$\widetilde{\mathcal{S}}$ indicate the variable-sets containing their
associated auxiliary variables. 
\begin{figure*}
\begin{eqnarray}
\underset{\widetilde{\mathcal{C}},\widetilde{\mathcal{P}},\widetilde{\mathcal{S}}}{\mbox{maximize}} &  & \sum_{i=1}^{N}\sum_{j\in\mathcal{G}_{n}}\left[\sum_{e_{1}\in\mathcal{E}_{n,j}}\widetilde{C}_{n,j,e_{1}}^{BU,T_{1}}+\sum_{e_{2}\in\mathcal{E}_{n,j}}\widetilde{C}_{n,j,e_{2}}^{BU,T_{2}}\right]+\left[\sum_{e\in\mathcal{E}_{n,j}}\widetilde{C}_{n,j,e}^{BRU}\right]\label{eq:obj_relax}\\
\mbox{subject to} &  & \frac{\widetilde{s}_{n,j}}{2}\log_{2}\left(1+\frac{G_{n,j,e_{1}}^{BU,T_{1}}\widetilde{P}_{n,j,e_{1}}^{BU,T_{1}}}{\widetilde{s}_{n,j}}\right)\geq\widetilde{C}_{n,j,e_{1}}^{BU,T_{1}}\mbox{, }\forall n,j,e_{1},\label{eq:phi_relax}\\
 &  & \frac{\widetilde{s}_{n,j}}{2}\log_{2}\left(1+\frac{G_{n,j,e_{2}}^{BU,T_{2}}\widetilde{P}_{n,j,e_{2}}^{BU,T_{2}}}{\widetilde{s}_{n,j}}\right)\geq\widetilde{C}_{n,j,e_{2}}^{BU,T_{2}}\mbox{, }\forall n,j,e_{2},\label{eq:varphi_relax}\\
 &  & \frac{\widetilde{s}_{n,j}}{2}\log_{2}\left(1+\frac{G_{n,j,e}^{BR,T_{1}}\widetilde{P}_{n,j,e}^{BR,T_{1}}}{\widetilde{s}_{n,j}}\right)\geq\widetilde{C}_{n,j,e}^{BRU}\mbox{, }\forall n,j,e,\label{eq:kappa_relax}\\
 &  & \frac{\widetilde{s}_{n,j}}{2}\log_{2}\left(1+\frac{G_{n,j,e}^{RU,T_{2}}\widetilde{P}_{n,j,e}^{RU,T_{2}}}{\widetilde{s}_{n,j}}\right)\geq\widetilde{C}_{n,j,e}^{BRU}\mbox{, }\forall n,j,e,\label{eq:varkappa_relax}\\
 &  & \sum_{j\in\mathcal{G}_{n}}\widetilde{s}_{n,j}\leq1\mbox{, }\forall n,\label{eq:theta_relax}\\
 &  & \sum_{i=1}^{N}\sum_{j\in\mathcal{G}_{n}}\left[\sum_{e_{1}\in\mathcal{E}_{n,j}}\widetilde{P}_{n,j,e_{1}}^{BU,T_{1}}+\sum_{e\in\mathcal{E}_{n,j}}\widetilde{P}_{n,j,e}^{BR,T_{1}}\right]\leq P_{max}^{B},\label{eq:zeta_relax}\\
 &  & \sum_{i=1}^{N}\sum_{j\in\mathcal{G}_{n}}\sum_{e_{2}\in\mathcal{E}_{n,j}}\widetilde{P}_{n,j,e_{2}}^{BU,T_{2}}\leq P_{max}^{B},\label{eq:omega_relax}\\
 &  & \sum_{i=1}^{N}\sum_{j\in\mathcal{G}_{n}}\sum_{\begin{subarray}{c}
e\in\mathcal{E}_{n,j}\\
\mathcal{M}\left(e\right)=m
\end{subarray}}\widetilde{P}_{n,j,e}^{RU,T_{2}}\leq P_{max}^{R}\mbox{, }\forall m,\label{eq:nu_relax}\\
 &  & 0\leq\widetilde{s}_{n,j}\leq1\mbox{, }\forall n,j,\label{eq:sigmatau_relax}\\
 &  & \widetilde{P}_{n,j,e_{1}}^{BU,T_{1}},\widetilde{P}_{n,j,e}^{BR,T_{1}},\widetilde{P}_{n,j,e_{2}}^{BU,T_{2}},\widetilde{P}_{n,j,e}^{RU,T_{2}}\geq0\mbox{, }\forall n,j,e_{1},e_{2},e.\label{eq:chipsi_relax}
\end{eqnarray}
\hrulefill{} 
\end{figure*}

It can be seen that the objective function of~(\ref{eq:obj_orig})
has been replaced by~(\ref{eq:obj_relax}) using the auxiliary rate
variables given in~(\ref{eq:aux_rates}), and by introducing the
hypograph constraints~(\ref{eq:phi_relax})--(\ref{eq:varkappa_relax})%
\footnote{Note that obtaining separate constraints for the first- and second-phase
power control variables associated with the relayed
transmission is made possible using the DF protocol. This then allowed
us to readily derive the optimal power control variables as the decoupled
water-filling solutions in Section~\ref{sub:OptPrimal}.%
}. These additional constraints ensure that the feasible
auxiliary rate variables do not exceed their counterparts calculated
on each link before using relaxation. As a result, the sum rate given
by~(\ref{eq:obj_relax}) invoking the feasible auxiliary rate variables
does not exceed the sum rate given by~(\ref{eq:C_T}) either.

As our next step, we prove that the problem described
by~(\ref{eq:obj_relax})--(\ref{eq:chipsi_relax}) is a concave
programming problem. Clearly,~(\ref{eq:obj_relax}) is affine,
hence concave, while~(\ref{eq:theta_relax})--(\ref{eq:chipsi_relax})
are all affine, and hence convex. Therefore, what remains is to show
that constraints~(\ref{eq:phi_relax})--(\ref{eq:varkappa_relax})
are convex as well. These remaining constraints may be written in
the form of 
\begin{equation}
C-\frac{s}{2}\log_{2}\left(1+\frac{GP}{s}\right)\leq0,\label{eq:rate_simple}
\end{equation}
where $s$, $P$ and $C$ are the decision variables. It may be readily
verified that $\left(1+GP\right)$ is affine and hence concave. Thus,
$\log_{2}\left(1+GP\right)$ is concave, since $\log_{2}\left(\cdot\right)$
is concave and non-decreasing as a function of its argument. The function
$s\log_{2}\left(1+\frac{GP}{s}\right)$ is a perspective transformation%
\footnote{Strictly speaking, the perspective transformation also requires that
$s>0$. However, convexity is also preserved for the situation when
$s=0$ as proven in~\cite{Ng2010}.%
}~\cite{Boyd2004} of $\log_{2}\left(1+GP\right)$, which preserves
concavity. Finally, $C-\frac{s}{2}\log_{2}\left(1+\frac{GP}{s}\right)$
is convex, since it is the sum of two convex functions. Since~(\ref{eq:rate_simple})
is convex, it is clear that constraints~(\ref{eq:phi_relax})--(\ref{eq:varkappa_relax})
are convex, and so~(\ref{eq:obj_relax})--(\ref{eq:chipsi_relax})
is a concave programming problem, whose solution
algorithm is presented in Section~\ref{sub:DualDecomp}.

\subsection{ESEM problem}

The ESE objective function, given by~(\ref{eq:eeff_obj}), is formed
by dividing the objective function~(\ref{eq:obj_relax}) by $P_{T}\left(\widetilde{\mathcal{P}},\widetilde{\mathcal{S}}\right)$,
which is obtained by substituting~(\ref{eq:aux_p1})--(\ref{eq:aux_p4})
into~(\ref{eq:P_T}) and introducing the relaxed variables $\widetilde{s}_{n,j}$.
\begin{figure*}
\begin{equation}
\frac{\sum\limits _{i=1}^{N}\sum\limits _{j\in\mathcal{G}_{n}}\left[\sum\limits _{e_{1}\in\mathcal{E}_{n,j}}\widetilde{C}_{n,j,e_{1}}^{BU,T_{1}}+\sum\limits _{e_{2}\in\mathcal{E}_{n,j}}\widetilde{C}_{n,j,e_{2}}^{BU,T_{2}}\right]+\left[\sum\limits _{e\in\mathcal{E}_{n,j}}\widetilde{C}_{n,j,e}^{BRU}\right]}{\left(P_{C}^{B}+M\cdot P_{C}^{R}\right)+\frac{1}{2}\sum\limits _{n=1}^{N}\sum\limits _{j\in\mathcal{G}_{n}}\left[\xi^{B}\left(\sum\limits _{e_{1}\in\mathcal{E}_{n,j}}\widetilde{P}_{n,j,e_{1}}^{BU,T_{1}}+\sum\limits _{e_{2}\in\mathcal{E}_{n,j}}\widetilde{P}_{n,j,e_{2}}^{BU,T_{2}}\right)+\sum\limits _{e\in\mathcal{E}_{n,j}}\left(\xi^{B}\widetilde{P}_{n,j,e}^{BR,T_{1}}+\xi^{R}\widetilde{P}_{n,j,e}^{RU,T_{2}}\right)\right]}\label{eq:eeff_obj}
\end{equation}
\hrulefill{} 
\end{figure*}

The objective function~(\ref{eq:eeff_obj}) is a linear-fractional
function, since it is a ratio of two affine functions. Thus the ESEM
problem can be solved using the Charnes-Cooper transformation of~\cite{Avriel1988},
as given by 
\begin{eqnarray}
\widehat{C}_{n,j,e_{1}}^{BU,T_{1}} & = & \widetilde{C}_{n,j,e_{1}}^{BU,T_{1}}t\mbox{, }\forall n,j,e_{1},\\
\widehat{C}_{n,j,e_{2}}^{BU,T_{2}} & = & \widetilde{C}_{n,j,e_{2}}^{BU,T_{2}}t\mbox{, }\forall n,j,e_{2},\\
\widehat{C}_{n,j,,e}^{BRU} & = & \widetilde{C}_{n,j,e}^{BRU}t\mbox{, }\forall n,j,e,\\
\widehat{P}_{n,j,e_{1}}^{BU,T_{1}} & = & \widetilde{P}_{n,j,e_{1}}^{BU,T_{1}}t\mbox{, }\forall n,j,e_{1},\\
\widehat{P}_{n,j,e_{2}}^{BU,T_{2}} & = & \widetilde{P}_{n,j,e_{2}}^{BU,T_{2}}t\mbox{, }\forall n,j,e_{2},\\
\widehat{P}_{n,j,e}^{BR,T_{1}} & = & \widetilde{P}_{n,j,e}^{BR,T_{1}}t\mbox{, }\forall n,j,e,\\
\widehat{P}_{n,j,e}^{RU,T_{2}} & = & \widetilde{P}_{n,j,e}^{RU,T_{2}}t\mbox{, }\forall n,j,e,\\
\widehat{s}_{n,j} & = & \widetilde{s}_{n,j}t\mbox{, }\forall n,j,
\end{eqnarray}
where the auxiliary variable $t$ is given by 
\begin{equation}
t=\frac{1}{P_{T}\left(\widetilde{\mathcal{P}},\widetilde{\mathcal{S}}\right)}.
\end{equation}
Thus, the ESEM problem may be written%
\footnote{Strictly speaking, the constraint $t>0$ is also needed, but this
is guaranteed due to constraint~(\ref{eq:varrho_eeff}).%
} as~(\ref{eq:obj_eeff})--(\ref{eq:varrho_eeff}), where $\widehat{\mathcal{C}}$,
$\widehat{\mathcal{P}}$ and $\widehat{\mathcal{S}}$ indicate the
variable-sets containing their associated transformed variables. It
is clear that the objective function~(\ref{eq:obj_eeff}) is affine,
hence concave, while the constraints~(\ref{eq:theta_eeff})--(\ref{eq:varrho_eeff})
are all affine, and hence convex. The constraints~(\ref{eq:phi_eeff})--(\ref{eq:varkappa_eeff})
are of the form~(\ref{eq:rate_simple}) and are hence convex. Therefore,
the problem described by~(\ref{eq:obj_eeff})--(\ref{eq:varrho_eeff})
is a concave programming problem, which can be solved using the algorithm
of Section~\ref{sub:DualDecomp}.

\begin{figure*}
\begin{eqnarray}
\underset{\widehat{\mathcal{C}},\widehat{\mathcal{P}},\widehat{\mathcal{S}},t}{\mbox{maximize}} &  & \sum_{i=1}^{N}\sum_{j\in\mathcal{G}_{n}}\left[\sum_{e_{1}\in\mathcal{E}_{n,j}}\widehat{C}_{n,j,e_{1}}^{BU,T_{1}}+\sum_{e_{2}\in\mathcal{E}_{n,j}}\widehat{C}_{n,j,e_{2}}^{BU,T_{2}}\right]+\left[\sum_{e\in\mathcal{E}_{n,j}}\widehat{C}_{n,j,e}^{BRU}\right]\label{eq:obj_eeff}\\
\mbox{subject to} &  & \frac{\widehat{s}_{n,j}}{2}\log_{2}\left(1+\frac{G_{n,j,e_{1}}^{BU,T_{1}}\widehat{P}_{n,j,e_{1}}^{BU,T_{1}}}{\widehat{s}_{n,j}}\right)\geq\widehat{C}_{n,j,e_{1}}^{BU,T_{1}}\mbox{, }\forall n,j,e_{1},\label{eq:phi_eeff}\\
 &  & \frac{\widehat{s}_{n,j}}{2}\log_{2}\left(1+\frac{G_{n,j,e_{2}}^{BU,T_{2}}\widehat{P}_{n,j,e_{2}}^{BU,T_{2}}}{\widehat{s}_{n,j}}\right)\geq\widehat{C}_{n,j,e_{2}}^{BU,T_{2}}\mbox{, }\forall n,j,e_{2},\label{eq:varphi_eeff}\\
 &  & \frac{\widehat{s}_{n,j}}{2}\log_{2}\left(1+\frac{G_{n,j,e}^{BR,T_{1}}\widehat{P}_{n,j,e}^{BR,T_{1}}}{\widehat{s}_{n,j}}\right)\geq\widehat{C}_{n,j,e}^{BRU}\mbox{, }\forall n,j,e,\label{eq:kappa_eeff}\\
 &  & \frac{\widehat{s}_{n,j}}{2}\log_{2}\left(1+\frac{G_{n,j,e}^{RU,T_{2}}\widehat{P}_{n,j,e}^{RU,T_{2}}}{\widehat{s}_{n,j}}\right)\geq\widehat{C}_{n,j,e}^{BRU}\mbox{, }\forall n,j,e,\label{eq:varkappa_eeff}\\
 &  & \sum_{j\in\mathcal{G}_{n}}\widehat{s}_{n,j}\leq t\mbox{, }\forall n,\label{eq:theta_eeff}\\
 &  & \sum_{i=1}^{N}\sum_{j\in\mathcal{G}_{n}}\left[\sum_{e_{1}\in\mathcal{E}_{n,j}}\widehat{P}_{n,j,e_{1}}^{BU,T_{1}}+\sum_{e\in\mathcal{E}_{n,j}}\widehat{P}_{n,j,e}^{BR,T_{1}}\right]\leq t\cdot P_{max}^{B},\label{eq:zeta_eeff}\\
 &  & \sum_{i=1}^{N}\sum_{j\in\mathcal{G}_{n}}\sum_{e_{2}\in\mathcal{E}_{n,j}}\widehat{P}_{n,j,e_{2}}^{BU,T_{2}}\leq t\cdot P_{max}^{B},\label{eq:omega_eeff}\\
 &  & \sum_{i=1}^{N}\sum_{j\in\mathcal{G}_{n}}\sum_{\begin{subarray}{c}
e\in\mathcal{E}_{n,j}\\
\mathcal{M}\left(e\right)=m
\end{subarray}}\widehat{P}_{n,j,e}^{RU,T_{2}}\leq t\cdot P_{max}^{R}\mbox{, }\forall m,\label{eq:nu_eeff}\\
 &  & 0\leq\widehat{s}_{n,j}\leq t\mbox{, }\forall n,j,\label{eq:sigmatau_eeff}\\
 &  & \widehat{P}_{n,j,e_{1}}^{BU,T_{1}},\widehat{P}_{n,j,e}^{BR,T_{1}},\widehat{P}_{n,j,e_{2}}^{BU,T_{2}},\widehat{P}_{n,j,e}^{RU,T_{2}}\geq0\mbox{, }\forall n,j,e_{1},e_{2},e,\label{eq:chipsi_eeff}\\
 &  & t\cdot\left(P_{C}^{B}+M\cdot P_{C}^{R}\right)\nonumber \\
 &  & +\frac{1}{2}\sum_{n=1}^{N}\sum_{j\in\mathcal{G}_{n}}\left[\xi^{B}\left(\sum_{e_{1}\in\mathcal{E}_{n,j}}\widehat{P}_{n,j,e_{1}}^{BU,T_{1}}+\sum_{e_{2}\in\mathcal{E}_{n,j}}\widehat{P}_{n,j,e_{2}}^{BU,T_{2}}\right)+\sum_{e\in\mathcal{E}_{n,j}}\left(\xi^{B}\widehat{P}_{n,j,e}^{BR,T_{1}}+\xi^{R}\widehat{P}_{n,j,e}^{RU,T_{2}}\right)\right]=1.\nonumber \\
\label{eq:varrho_eeff}
\end{eqnarray}
\hrulefill{} 
\end{figure*}

\subsection{Dual decomposition based solution algorithm}\label{sub:DualDecomp}

The dual decomposition method of~\cite{Palomar2006,Cheung2013} may
be used for conceiving solution algorithms for our
SEM and ESEM problems formulated as~(\ref{eq:obj_relax})--(\ref{eq:chipsi_relax})
and~(\ref{eq:obj_eeff})--(\ref{eq:varrho_eeff}), respectively.
We commence by describing the solution algorithm conceived
for~(\ref{eq:obj_eeff})--(\ref{eq:varrho_eeff}), which we term
the ESEM algorithm. The ESEM algorithm, based on dual decomposition,
iterates between calculating the tentative optima
of the primal variables, namely $\widehat{C}_{n,j,e_{1}}^{BU,T_{1}}$,
$\widehat{C}_{n,j,e_{2}}^{BU,T_{2}}$, $\widehat{C}_{n,j,e}^{BRU}$,
$\widehat{P}_{n,j,e_{1}}^{BU,T_{1}}$, $\widehat{P}_{n,j,e_{2}}^{BU,T_{2}}$,
$\widehat{P}_{n,j,e}^{BR,T_{1}}$, $\widehat{P}_{n,j,e}^{RU,T_{2}}$,
$\widehat{s}_{n,j}$ as well as $t$, and updating
the dual variables $\lambda^{T_{1}}$, $\lambda^{T_{2}}$, $\nu_{m}$
as well as $\mu$, which will be defined later, until the objective
function value converges.

\subsubsection{Calculating tentative optima of primal variables\label{sub:OptPrimal}}

Based on our previous work~\cite{Cheung2013} that
employed the dual decomposition and by employing the
Karush-Kuhn-Tucker optimality conditions~\cite{Boyd2004}, we reveal
that the tentatively optimal transformed power control variables for the direct
SMCs encountered in the problem of~(\ref{eq:obj_eeff})--(\ref{eq:varrho_eeff})
may be formulated as the water-filling solutions of%
\footnote{In this paper, $\left[\cdot\right]^{+}$ is equivalent to $\max\left(0,\cdot\right)$.%
} 
\begin{eqnarray}
\widehat{P}_{n,j,e_{1}}^{BU,T_{1}} & = & \widehat{s}_{n,j}\left[\frac{1}{\left(\xi^{B}\mu+2\lambda^{T_{1}}\right)\ln2}-\frac{1}{G_{n,j,e_{1}}^{BU,T_{1}}}\right]^{+}\nonumber \\
 & = & \widehat{s}_{n,j}P_{n,j,e_{1}}^{BU,T_{1}}\label{eq:optimum_direct_SMC_power_T1}
\end{eqnarray}
and 
\begin{eqnarray}
\widehat{P}_{n,j,e_{2}}^{BU,T_{2}} & = & \widehat{s}_{n,j}^ {}\left[\frac{1}{\left(\xi^{B}\mu+2\lambda^{T_{2}}\right)\ln2}-\frac{1}{G_{n,j,e_{2}}^{BU,T_{2}}}\right]^{+}\nonumber \\
 & = & \widehat{s}_{n,j}P_{n,j,e_{2}}^{BU,T_{2}}.\label{eq:optimum_direct_SMC_power_T2}
\end{eqnarray}
In addition, the transformed power control variables for the relaying
SMCs may be \textit{initially} written
as 
\begin{eqnarray}
\widehat{P}_{n,j,e}^{BR,T_{1}} & = & \widehat{s}_{n,j}^ {}\left[\frac{1}{\left(\xi^{B}\mu+2\lambda^{T_{1}}\right)\ln2}-\frac{1}{G_{n,j,e}^{BR,T_{1}}}\right]^{+}\nonumber \\
 & = & \widehat{s}_{n,j}P_{n,j,e}^{BR,T_{1}}\label{eq:optimum_relay_SMC_power_T1}
\end{eqnarray}
and 
\begin{eqnarray}
\widehat{P}_{n,j,e}^{RU,T_{2}} & = & \widehat{s}_{n,j}^ {}\left[\frac{1}{\left(\xi^{R}\mu+2\nu_{\mathcal{M}\left(e\right)}\right)\ln2}-\frac{1}{G_{n,j,e}^{RU,T_{2}}}\right]^{+}\nonumber \\
 & = & \widehat{s}_{n,j}P_{n,j,e}^{RU,T_{2}}.\label{eq:optimum_relay_SMC_power_T2}
\end{eqnarray}
Note that the value of $\widehat{s}_{n,j}$ in~(\ref{eq:optimum_direct_SMC_power_T1})--(\ref{eq:optimum_relay_SMC_power_T2})
is not yet known. Since the SE attainable for a relaying
link is limited by the weaker of the BS-to-RN and RN-to-UE links,
there is no need to transmit at a high power on
the stronger link, if the other link is unable to support the high
SE. Thus, the tentatively optimal transformed power control variables provided
for the relaying SMC $e$ may be refined by substituting~(\ref{eq:optimum_relay_SMC_power_T1})--(\ref{eq:optimum_relay_SMC_power_T2})
into the right-hand side of 
\begin{equation}
\widehat{P}_{n,j,e}^{BR,T_{1}}=\min\left(\widehat{P}_{n,j,e}^{BR,T_{1}},\frac{G_{n,j,e}^{RU,T_{2}}}{G_{n,j,e}^{BR,T_{1}}}\cdot\widehat{P}_{n,j,e}^{RU,T_{2}}\right)
\end{equation}
and 
\begin{equation}
\widehat{P}_{n,j,e}^{RU,T_{2}}=\min\left(\widehat{P}_{n,j,e}^{RU,T_{2}},\frac{G_{n,j,e}^{BR,T_{1}}}{G_{n,j,e}^{RU,T_{2}}}\cdot\widehat{P}_{n,j,e}^{BR,T_{1}}\right).
\end{equation}

As a result, the tentative estimates of the
maximum values that $\widehat{C}_{n,j,e_{1}}^{BU,T_{1}}$, $\widehat{C}_{n,j,e_{2}}^{BU,T_{2}}$
and $\widehat{C}_{n,j,e}^{BRU}$ can attain are given by 
\begin{equation}
\widehat{C}_{n,j,e_{1}}^{BU,T_{1}}=\frac{\widehat{s}_{n,j}}{2}\log_{2}\left(1+\frac{G_{n,j,e_{1}}^{BU,T_{1}}\widehat{P}_{n,j,e_{1}}^{BU,T_{1}}}{\widehat{s}_{n,j}}\right),
\end{equation}
\begin{equation}
\widehat{C}_{n,j,e_{2}}^{BU,T_{2}}=\frac{\widehat{s}_{n,j}}{2}\log_{2}\left(1+\frac{G_{n,j,e_{2}}^{BU,T_{2}}\widehat{P}_{n,j,e_{2}}^{BU,T_{2}}}{\widehat{s}_{n,j}}\right),
\end{equation}
and 
\begin{eqnarray}
\widehat{C}_{n,j,e}^{BRU} & = & \frac{\widehat{s}_{n,j}}{2}\log_{2}\left(1+\frac{G_{n,j,e}^{BR,T_{1}}\widehat{P}_{n,j,e}^{BR,T_{1}}}{\widehat{s}_{n,j}}\right)\nonumber \\
 & = & \frac{\widehat{s}_{n,j}}{2}\log_{2}\left(1+\frac{G_{n,j,e}^{RU,T_{2}}\widehat{P}_{n,j,e}^{RU,T_{2}}}{\widehat{s}_{n,j}}\right),\nonumber \\
\end{eqnarray}
where the value of $\widehat{s}_{n,j}$ remains
unknown. However, it is plausible that for the
purpose of maximizing the objective function value, $\widehat{s}_{n,j},\forall n,j$
will always be given its maximum value $t$, if the single SMC group
$j$ is selected for subcarrier block $n$. Thus, the tentatively optimal SMC
group $j$ for subcarrier block $n$ is given by the group obtaining
the highest value of 
\begin{equation}
\sum_{j\in\mathcal{G}_{n}}\left[\sum_{e_{1}\in\mathcal{E}_{n,j}}\widehat{C}_{n,j,e_{1}}^{BU,T_{1}}+\sum_{e_{2}\in\mathcal{E}_{n,j}}\widehat{C}_{n,j,e_{2}}^{BU,T_{2}}\right]+\sum_{e\in\mathcal{E}_{n,j}}\widehat{C}_{n,j,e}^{BRU}.\label{eq:part_obj_to_max}
\end{equation}
where $\widehat{s}_{n,j}$ inside the logarithm functions may be canceled
out. Additionally, we can ignore the common positive multiplicative
factor of $t$ without affecting the maximization of (\ref{eq:part_obj_to_max}).
The objective function~(\ref{eq:obj_eeff}) is maximized when choosing
this particular group $j$ for subcarrier block $n$, while for the
remaining groups associated with the same subcarrier block, we set
$\widehat{P}_{n,j'\ne j,e_{1}}^{BU,T_{1}}=\widehat{P}_{n,j'\ne j,e_{2}}^{BU,T_{2}}=\widehat{P}_{n,j'\ne j,e}^{BR,T_{1}}=\widehat{P}_{n,j'\ne j,e}^{RU,T_{2}}=\widehat{s}_{n,j'\ne j}^ {}=\widehat{C}_{n,j'\ne j,e_{1}}^{BU,T_{1}}=\widehat{C}_{n,j'\ne j,e_{2}}^{BU,T_{2}}=\widehat{C}_{n,j'\ne j,e}^{BRU}=P_{n,j'\ne j,e_{1}}^{BU,T_{1}}=P_{n,j'\ne j,e_{2}}^{BU,T_{2}}=P_{n,j'\ne j,e}^{BR,T_{1}}=P_{n,j'\ne j,e}^{RU,T_{2}}=0$,
as these remaining groups are not chosen.

\begin{figure*}
\begin{equation}
t=\left(P_{C}^{B}+M\cdot P_{C}^{R}+\frac{1}{2}\sum_{n=1}^{N}\sum_{j\in\mathcal{G}_{n}}\left[\xi^{B}\left(\sum_{e_{1}\in\mathcal{E}_{n,j}}P_{n,j,e_{1}}^{BU,T_{1}}+\sum_{e_{2}\in\mathcal{E}_{n,j}}P_{n,j,e_{2}}^{BU,T_{2}}\right)+\sum_{e\in\mathcal{E}_{n,j}}\left(\xi^{B}P_{n,j,e}^{BR,T_{1}}+\xi^{R}P_{n,j,e}^{RU,T_{2}}\right)\right]\right)^{-1}\label{eq:optT}
\end{equation}
\hrulefill{} 
\end{figure*}

Consequently, the value of $t$ is given by~(\ref{eq:optT}). Note that this is possible without knowing the exact
value of $\widehat{s}_{n,j}$,  since the factor
of $\widehat{s}_{n,j}$ may be canceled out, and thus~(\ref{eq:optT})
is only dependent on the dual variables and on the tentatively optimal SMC group
selection.

Having identified the tentative optimal SMC group, we
set $\widehat{s}_{n,j}=t$ for this selected SMC group corresponding
to each subcarrier block $n$, and we have
\begin{equation}
\widehat{C}_{n,j,e_{1}}^{BU,T_{1}}=\frac{t}{2}\log_{2}\left(1+G_{n,j,e_{1}}^{BU,T_{1}}P_{n,j,e_{1}}^{BU,T_{1}}\right),
\end{equation}
\begin{equation}
\widehat{C}_{n,j,e_{2}}^{BU,T_{2}}=\frac{t}{2}\log_{2}\left(1+G_{n,j,e_{2}}^{BU,T_{2}}P_{n,j,e_{2}}^{BU,T_{2}}\right)
\end{equation}
as well as
\begin{eqnarray}
\widehat{C}_{n,j,e}^{BRU} & = & \frac{t}{2}\log_{2}\left(1+G_{n,j,e}^{BR,T_{1}}P_{n,j,e}^{BR,T_{1}}\right)\nonumber \\
 & = & \frac{t}{2}\log_{2}\left(1+G_{n,j,e}^{RU,T_{2}}P_{n,j,e}^{RU,T_{2}}\right),\nonumber \\
\end{eqnarray}
for that selected SMC group. To summarize, given a set of dual variables,
the values of power control variables are obtained, resulting in an
tentatively optimal SMC group, which obtains the SE values for
the corresponding subcarrier block. Therefore, all of the
primal variables are obtained for a given set of dual variables. Thus, they are jointly optimized.

\subsubsection{Updating the dual variables\label{sub:UpdateDual}}

From the derivation of the optimal primal variables described in Section~\ref{sub:OptPrimal},
we can see that the constraints~(\ref{eq:phi_eeff})--(\ref{eq:theta_eeff})
and~(\ref{eq:sigmatau_eeff})--(\ref{eq:varrho_eeff}) are implicitly
satisfied. Therefore, we update the dual variables
$\lambda^{T_{1}}$, $\lambda^{T_{2}}$ and $\nu_{m}$ which are associated
with the remaining constraints~(\ref{eq:zeta_eeff})--(\ref{eq:nu_eeff}),
respectively. These may be viewed as pricing parameters to ensure
that the optimal power control variables satisfy~(\ref{eq:zeta_eeff})--(\ref{eq:nu_eeff}).
\begin{figure*}
\begin{equation}
\lambda^{T_{1}}\left(i\right)=\left[\lambda^{T_{1}}\left(i-1\right)-\delta_{\lambda^{T_{1}}}\left(i\right)\cdot\left(t\cdot P_{max}^{B}-\sum_{i=1}^{N}\sum_{j\in\mathcal{G}_{n}}\left[\sum_{e_{1}\in\mathcal{E}_{n,j}}\widehat{P}_{n,j,e_{1}}^{BU,T_{1}}+\sum_{e\in\mathcal{E}_{n,j}}\widehat{P}_{n,j,e}^{BR,T_{1}}\right]\right)\right]^{+}\label{eq:lambda1_up}
\end{equation}

\begin{equation}
\lambda^{T_{2}}\left(i\right)=\left[\lambda^{T_{2}}\left(i-1\right)-\delta_{\lambda^{T_{2}}}\left(i\right)\cdot\left(t\cdot P_{max}^{B}-\sum_{i=1}^{N}\sum_{j\in\mathcal{G}_{n}}\sum_{e_{2}\in\mathcal{E}_{n,j}}\widehat{P}_{n,j,e_{2}}^{BU,T_{2}}\right)\right]^{+}\label{eq:lambda2_up}
\end{equation}

\begin{equation}
\nu_{m}\left(i\right)=\left[\nu_{m}\left(i-1\right)-\delta_{\nu_{m}}\left(i\right)\cdot\left(t\cdot P_{max}^{R}-\sum_{i=1}^{N}\sum_{j\in\mathcal{G}_{n}}\sum_{\begin{subarray}{c}
e\in\mathcal{E}_{n,j}\\
\mathcal{M}\left(e\right)=m
\end{subarray}}\widehat{P}_{n,j,e}^{RU,T_{2}}\right)\right]^{+}\mbox{, }\forall m\label{eq:nu_up}
\end{equation}
 
\begin{eqnarray}
\mu\left(i\right) & = & t\cdot\Bigg(\sum_{i=1}^{N}\sum_{j\in\mathcal{G}_{n}}\left[\sum_{e_{1}\in\mathcal{E}_{n,j}}\widetilde{C}_{n,j,e_{1}}^{BU,T_{1}}+\sum_{e_{2}\in\mathcal{E}_{n,j}}\widetilde{C}_{n,j,e_{2}}^{BU,T_{2}}\right]+\left[\sum_{e\in\mathcal{E}_{n,j}}\widetilde{C}_{n,j,e}^{BRU}\right]\nonumber \\
 &  & +\lambda^{T_{1}}\left(i-1\right)\cdot\left(P_{max}^{B}-\sum_{i=1}^{N}\sum_{j\in\mathcal{G}_{n}}\left[\sum_{e_{1}\in\mathcal{E}_{n,j}}\widetilde{P}_{n,j,e_{1}}^{BU,T_{1}}+\sum_{e\in\mathcal{E}_{n,j}}\widetilde{P}_{n,j,e}^{BR,T_{1}}\right]\right)\nonumber \\
 &  & +\lambda^{T_{2}}\left(i-1\right)\cdot\left(P_{max}^{B}-\sum_{i=1}^{N}\sum_{j\in\mathcal{G}_{n}}\sum_{e_{2}\in\mathcal{E}_{n,j}}\widetilde{P}_{n,j,e_{2}}^{BU,T_{2}}\right)\nonumber \\
 &  & +\sum_{m=1}^{M}\nu_{m}\left(i-1\right)\cdot\left(P_{max}^{R}-\sum_{i=1}^{N}\sum_{j\in\mathcal{G}_{n}}\sum_{\begin{subarray}{c}
e\in\mathcal{E}_{n,j}\\
\mathcal{M}\left(e\right)=m
\end{subarray}}\widetilde{P}_{n,j,e}^{RU,T_{2}}\right)\Bigg)\label{eq:mu_up}
\end{eqnarray}
 \hrulefill{} 
\end{figure*}

Since the Lagrangian of~(\ref{eq:obj_eeff})--(\ref{eq:varrho_eeff})
is differentiable w.r.t. the dual variables, at each iteration $i$
of the solution algorithm, these dual variables may be updated according
to~(\ref{eq:lambda1_up})--(\ref{eq:nu_up}), where $\delta_{\lambda^{T_{1}}}\left(i\right)$,
$\delta_{\lambda^{T_{2}}}\left(i\right)$ and $\delta_{\nu_{m}}\left(i\right)$
are appropriately chosen step sizes~\cite{Palomar2006} at iteration
$i$.

The remaining dual variable, $\mu$, which is associated
with~(\ref{eq:varrho_eeff}) must also be updated. However, the constraint
given by~(\ref{eq:varrho_eeff}) is implicitly satisfied since the
value of $t$ is computed from~(\ref{eq:optT}). Therefore, we opt
for an alternative method based on differentiating the Lagrangian
w.r.t $t$ and substituting in the intermediate values of $\widehat{\mathcal{C}}$,
$\widehat{\mathcal{P}}$, $\widehat{\mathcal{S}}$ and $t$. Thus,
the updated value of $\mu$ is given by~(\ref{eq:mu_up}). 

All primal variables are jointly optimized in Section~\ref{sub:OptPrimal}
as the optimal power variables are determined by the related
dual variables. This leads to the optimal group
selection and rate variables, which then allow us
to find the optimal $t$. Given the tentative optima
of primal variables, the algorithm proceeds to update the dual variables,
which are mostly to ensure that the maximum power
constraints are not violated. Using these updated dual variables,
the algorithm repeats this process until the objective function value
$\widehat{\eta}_{E}\left(i\right)$ at iteration $i$ reaches the
predefined convergence threshold, which is given
by $\left|\widehat{\eta}_{E}\left(i\right)-\widehat{\eta}_{E}\left(i-1\right)\right|<\epsilon$.

The method presented in Section~\ref{sub:OptPrimal} and Section~\ref{sub:UpdateDual}
solves the ESEM problem described by~(\ref{eq:obj_eeff})--(\ref{eq:varrho_eeff}).
It may also be invoked for solving the SEM problem
of~(\ref{eq:obj_relax})--(\ref{eq:chipsi_relax}), while fixing
$\mu=0$ and $t=1$. This is because the ESEM problem considered is
simplified to the SEM problem, when we have $\mu=0$ and $t=1$ .

\section{Numerical results and discussions\label{sec:ResDis}}

\begin{table}[t]
\centering{}\protect\caption{Simulation parameters used to obtain all results in Section~\ref{sec:ResDis}
unless otherwise specified.}

\label{tab:param} \setlength{\extrarowheight}{1.5pt} %
\begin{tabular}{|l|r|}
\hline 
\textbf{Simulation parameter}  & \textbf{Value}\tabularnewline
\hline 
\hline 
Subcarrier block bandwidth, $W$ {[}Hertz{]}  & $180$k\tabularnewline
\hline 
Number of RNs, $M$  & $\{0,1,2,4\}$\tabularnewline
\hline 
Number of subcarriers blocks, $N$  & $\{6,12,25,50,100\}$\tabularnewline
\hline 
Number of UEs, $K$  & $\{2,10\}$\tabularnewline
\hline 
Antenna configuration, $\left(N_{B},N_{R},N_{U}\right)$  & $\left(4,4,2\right)$\tabularnewline
\hline 
Cell radius, {[}km{]}  & $\{0.75,1.25,1.75,2.25\}$\tabularnewline
\hline 
Ratio of BS-to-RN distance to the cell  & $0.5$\tabularnewline
radius, $D_{r}$  & \tabularnewline
\hline 
SNR gap of wireless transceivers, $\Delta\gamma$ {[}dB{]}  & 0\tabularnewline
\hline 
\multicolumn{1}{|l|}{Maximum total transmission power of the} & $\{0,10,20,30,$\tabularnewline
BS and RNs, $P_{max}^{B}$ and $P_{max}^{R}$ {[}dBm{]}  & $40,50,60\}$\tabularnewline
\hline 
Fixed power rating of the BS,  & 32.306$N_{B}$\tabularnewline
$P_{C}^{(B)}$ {[}Watts{]}~\cite{Arnold2010,Auer2010}  & \tabularnewline
\hline 
Fixed power rating of RNs,  & 21.874$N_{R}$\tabularnewline
$P_{C}^{(R)}$ {[}Watts{]}~\cite{Arnold2010,Auer2010}  & \tabularnewline
\hline 
Reciprocal of the BS power amplifier's  & 3.24$N_{B}$\tabularnewline
drain efficiency, $\xi^{(B)}$~\cite{Arnold2010,Auer2010}  & \tabularnewline
\hline 
Reciprocal of the RNs' power amplifier's  & 4.04$N_{R}$\tabularnewline
drain efficiency, $\xi^{(R)}$~\cite{Arnold2010,Auer2010}  & \tabularnewline
\hline 
Noise power spectral density, $N_{0}$ {[}dBm/Hz{]}  & \textminus 174\tabularnewline
\hline 
Convergence threshold, $\epsilon$  & $10^{-6}$\tabularnewline
\hline 
Number of channel samples  & $10^{4}$\tabularnewline
\hline 
\end{tabular}
\end{table}

This section presents the numerical results obtained, when employing
the SEM and ESEM algorithms%
\footnote{In all cases, the step sizes and the initial values of the dual variables
described in Section~\ref{sub:UpdateDual} are empirically optimized
to give the optimal objective function value in as
few iterations as possible, although the exact analytical method for
determining the optimal step sizes and initial values
still remains an open issue. In our experience, the algorithms converge
within just $10$ iterations when carefully chosen step sizes are
employed, regardless of the size of the problem.%
} described in Section~\ref{sec:ProbForm} to the MIMO-OFDMA multi-relay
cellular network considered. The pertinent simulation parameters are
given in Table~\ref{tab:param}. Additionally, the path-loss effect
is characterized relying on the method and parameters
of~\cite{3GPP_PL}, where the BS-to-UE and RN-to-UE
links are assumed to be non-line-of-sight~(NLOS) links, since these
links are typically blocked by buildings and other large obstructing
objects, while the BS-to-RN links are realistically
assumed to be line-of-sight~(LOS) links, as the RNs may be strategically
deployed on tall buildings to create strong wireless backhaul links.
Furthermore, independently and randomly generated set of UE locations
as well as fading channel realizations were used for each channel
sample.

The results of a baseline algorithm is also presented to highlight
the improved performance obtained from employing the SEM and ESEM
algorithms. This baseline algorithm consists of a random SMC grouping~(RG)
selection for each subcarrier block and then equal power allocation~(EPA)
across all the selected SMCs, and will be termed the RG-EPA
algorithm.

\subsection{On the optimality and the relative complexity of ESGA and OCGA for
various $\alpha$ values}

\begin{figure}
\begin{centering}
\includegraphics[scale=0.9]{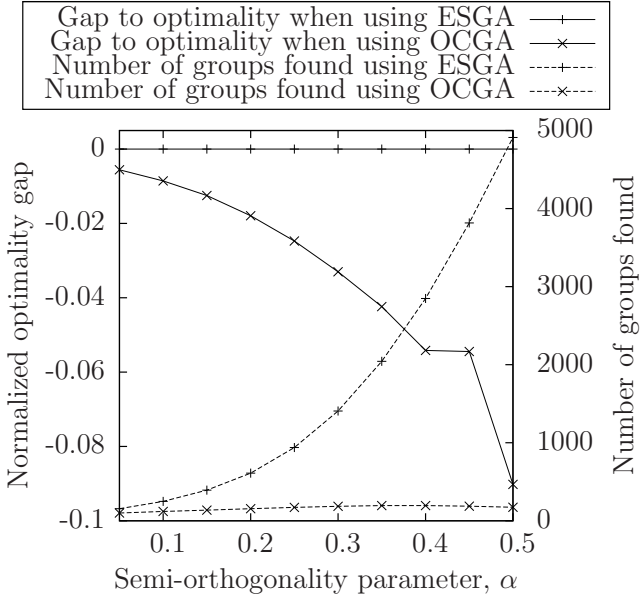} 
\par\end{centering}

\protect\caption{The optimality gap and total number of SMC groups
found when employing the ESGA and OCGA, and using the
parameters in Table~\ref{tab:param} with $N=6$, $K=2$, $M=2$,
$P_{max}^{B}=20$dBm, $P_{max}^{R}=10$dBm and a cell radius of $0.75$km.}

\label{fig:complexity} 
\end{figure}

Firstly, the behavior of the ESGA
and OCGA as a function of $\alpha$ is examined. Note in Fig.~\ref{fig:complexity}
that since the ESGA is capable of
enumerating all possible SMC groupings, which satisfy~(\ref{eq:semi-ortho})
for the corresponding $\alpha$, the optimal SE is attained. The 'normalized
optimality gap' is then defined as $\left(\beta/\beta^{*}\right)-1$,
where $\beta^{*}$ is the optimal SE obtained from employing the ESGA
algorithm, and $\beta$ is the SE obtained from any
other algorithm. We can see from Fig.~\ref{fig:complexity},
that the normalized optimality gap of OCGA relative to ESGA is about
$-0.005\sim-0.1$ for the $\alpha$ values considered.
However, the number of groups found using ESGA is exponentially increasing
with $\alpha$. By contrast, for OCGA, this number
is always significantly lower and gradually becomes less than $200$,
when $\alpha$ increases to $0.5$. In fact, the number of groups
found by OCGA is reduced to about $3.5\%$ of that found by ESGA
at $\alpha=0.5$. This demonstrates the viability of using OCGA in
the following simulations as a reduced-complexity near-optimum alternative
to ESGA. Under the conditions considered in Fig.~\ref{fig:complexity},
the optimal ESE solution is the same as the optimal SE solution, as
detailed in the next subsection. Therefore, as far
as ESEM is concerned, similar conclusions may be drawn
regarding the optimality of the two grouping algorithms.

\subsection{The variation in achievable SE and ESE for different values of $P_{max}^{B}$
and $P_{max}^{R}$}

\begin{figure}
\begin{centering}
\subfloat[Surface plots of the achievable SE when using the SEM, ESEM and RG-EPA
algorithms.]{\begin{centering}
\includegraphics[scale=0.7]{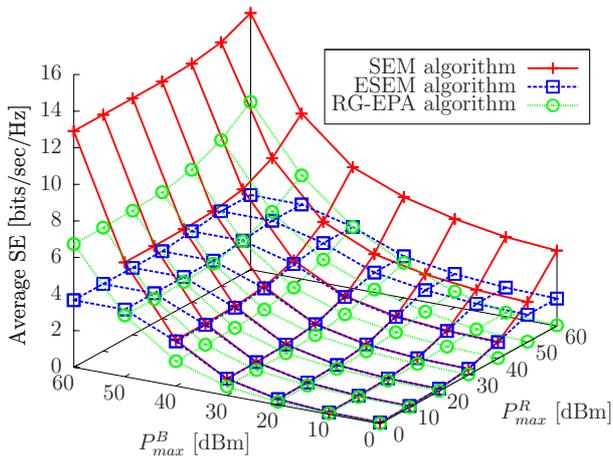} 
\par\end{centering}

}
\par\end{centering}

\begin{centering}
\subfloat[Surface plots of the achievable ESE when using the SEM, ESEM and RG-EPA
algorithms.]{\begin{centering}
\includegraphics[scale=0.7]{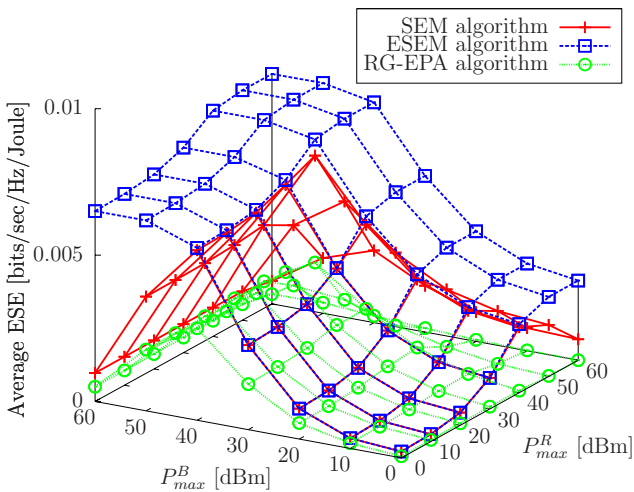} 
\par\end{centering}

}
\par\end{centering}

\protect\caption{The average achievable SE and ESE of the SEM, ESEM and RG-EPA algorithms
upon varying $P_{max}^{B}$ and $P_{max}^{R}$. The
parameters in Table~\ref{tab:param} with $N=6$, $K=10$, $M=2$,
$\alpha=0.1$ and a cell radius of $1.75$km are used.}

\label{fig:powers} 
\end{figure}

As shown in Fig.~\ref{fig:powers}(a), the achievable SE is monotonically
increasing with $P_{max}^{B}$ and $P_{max}^{R}$ when using the SEM
algorithm. This is not unexpected, since the SEM algorithm optimally
allocates all the available power for the sake of achieving the maximum
SE. By comparison, we observed from Fig.~\ref{fig:powers}(a)
and Fig.~\ref{fig:powers}(b) that both the achievable SE and ESE
of the ESEM algorithm saturate at some moderate values of $P_{max}^{B}$
and/or $P_{max}^{R}$. This is because the ESEM algorithm only allocates
``just'' enough power (that may be lower than the power budget values
of $P_{max}^{B}$ and/or $P_{max}^{R}$) for the sake of achieving
the maximum ESE. On the other hand, the ESE performance
of the SEM algorithm is severely degraded upon further increasing
$P_{max}^{B}$ and/or $P_{max}^{R}$ after its ESE performance reaches
the peak, as shown in Fig.~\ref{fig:powers}(b). This
is because the ESE metric is a quasiconcave function of the transmit
powers -- its numerator (i.e. the SE) increases logarithmically with
the transmit powers, while its denominator increases linearly with
the transmit powers. In fact, the peak ESE of the SEM algorithm is
attained at $P_{max}^{B}=40$dBm and $P_{max}^{R}=40$dBm, as seen
in Fig.~\ref{fig:powers}(b), and the associated normalized optimality
gap is only $-0.074$. By contrast, the achievable ESE when using
the ESEM algorithm also saturates at around $P_{max}^{B}=40$dBm and
$P_{max}^{R}=40$dBm%
\footnote{Note that when $P_{max}^{B}$ and $P_{max}^{R}$ have low/moderate
values, the SEM and ESEM algorithms share the same solutions of $\mathcal{P}$
and $\mathcal{S}$.%
}. Thus, the operating point of ``$P_{max}^{B}=40$dBm and $P_{max}^{R}=40$dBm''
may strike an attractive balance between SEM and ESEM.
Of course, the required trade-off may be struck on a case-by-case
basis in practical systems.

Additionally, the RG-EPA algorithm performs significantly worse in
terms of SE when compared to the SEM algorithm, and in terms of ESE
when compared to the ESEM algorithm. Furthermore, the RG-EPA algorithm
performs even worse than the SEM algorithm in terms of ESE. Although
the obtained SE when using the RG-EPA algorithm is, in some cases,
higher than the SE obtained when using the ESE algorithm, this performance
improvement comes at a great cost to the ESE performance of the RG-EPA
algorithm.

Finally, note that although both the SE of the SEM algorithm, and
the ESE of the ESEM algorithm are non-decreasing as either $P_{max}^{B}$
or $P_{max}^{R}$ is increased, the effect of increasing $P_{max}^{B}$
on the SE or ESE is significantly more pronounced, than that of applying
the same increase to $P_{max}^{R}$. The intuitive reasoning behind
this is that the power available at the BS has a more pronounced effect
on the system's performance, since the direct links and, more importantly,
the BS-to-RN links rely on the BS. Therefore, increasing $P_{max}^{R}$
is futile if the BS-to-RN links are not allocated sufficient power
to support the RN-to-UE links.

\subsection{The achievable SE and ESE as a function of $M$ and the cell radius}

\begin{figure}
\begin{centering}
\subfloat[Surface plots of the achievable SE when using the SEM, ESEM and RG-EPA
algorithms.]{\begin{centering}
\includegraphics[scale=0.7]{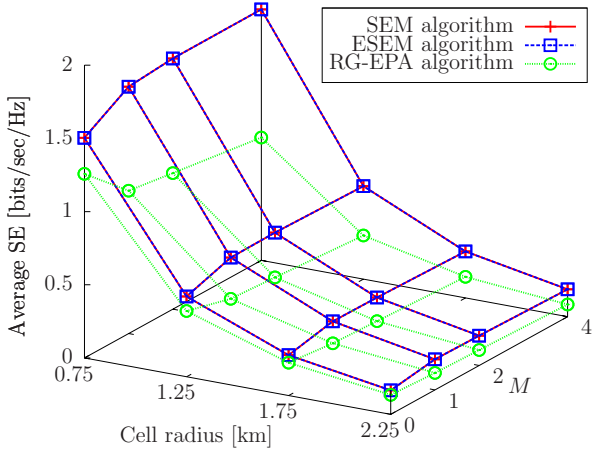} 
\par\end{centering}

}
\par\end{centering}

\begin{centering}
\subfloat[Surface plots of the achievable ESE when using the SEM, ESEM and RG-EPA
algorithms.]{\begin{centering}
\includegraphics[scale=0.7]{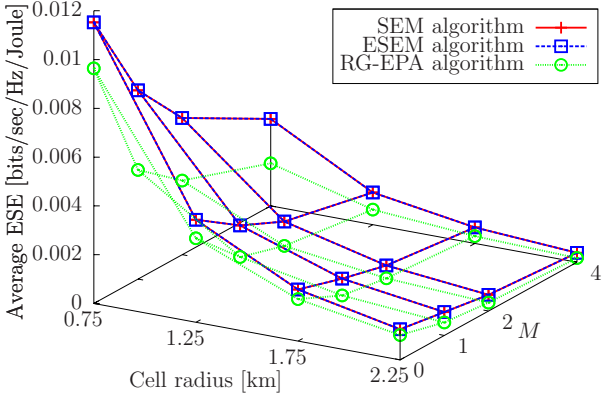} 
\par\end{centering}

}
\par\end{centering}

\protect\caption{The average achievable SE and ESE of the SEM, ESEM and RG-EPA algorithms
upon varying $M$ and cell radius, and using the parameters in Table~\ref{tab:param}
with $N=6$, $K=10$, $\alpha=0.1$, $P_{max}^{B}=20$dBm and $P_{max}^{R}=10$dBm.}

\label{fig:relays} 
\end{figure}

Fig.~\ref{fig:relays} illustrates some advantages and disadvantages
of employing RNs in the cellular system considered. We
observe that the specific low values of the power
constraints result in the same solutions for both the SEM and ESEM
algorithms. This phenomenon was also shown in Fig.~\ref{fig:powers}.

As evidenced in Fig.~\ref{fig:relays}(a), the attainable SE increases
with $M$, which is a benefit of the additional selection diversity,
when forming relaying links. However, the attainable SE does not increase
substantially beyond $M=2$. In fact, only an increase of $0.1\%$
is attained for the SE when $M$ is increased from $2$ to $4$ at
a cell radius of $0.75$km. On the other hand, the cost in terms of
ESE is significant~($36.4\%$), as shown in Fig.~\ref{fig:relays}(b).
This suggests that employing RNs does not constitute an energy-spectral-efficient
technique although it increases the SE of a cellular
system, which is partially due to the power amplifier inefficiency
and owing to the non-negligible fixed circuit energy dissipation.
Note furthermore that both the attainable SE and ESE are decreasing
upon increasing the cell radius as a result of the increased path-loss
of all the wireless links. However, this reduction is relatively small
between a cell radius of $1.75$km and $2.25$km. The reason behind
this phenomenon is that both the SEM and ESEM algorithms will selectively
serve the UEs nearer to the BS, so that a similar performance may
be attained without suffering from a substantial path-loss. This is
also the reason why the gain in SE gleaned by employing
RNs at a cell radius of $2.25$km seems negligible in Fig.~\ref{fig:relays}(a).
Once again, the RG-EPA algorithm performs worse both in terms of SE
and ESE performance.

\subsection{The achievable SE and ESE as a function of $N$ and $N_{B}$}

\begin{figure}
\begin{centering}
\subfloat[Surface plots of the achievable SE when using the SEM, ESEM and RG-EPA
algorithms.]{\begin{centering}
\includegraphics[scale=0.7]{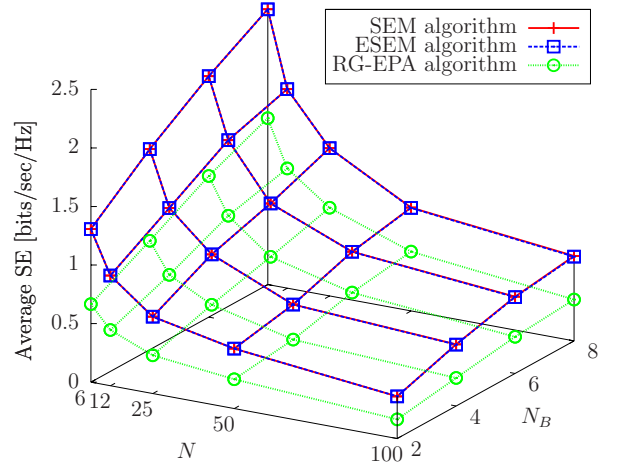} 
\par\end{centering}

}
\par\end{centering}

\begin{centering}
\subfloat[Surface plots of the achievable ESE when using the SEM, ESEM and RG-EPA
algorithms.]{\begin{centering}
\includegraphics[scale=0.7]{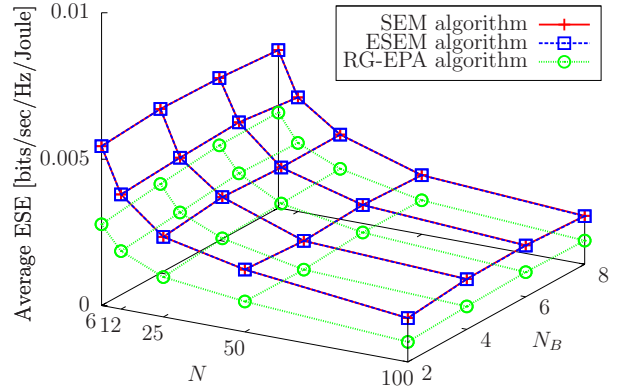} 
\par\end{centering}

}
\par\end{centering}

\protect\caption{The average achievable SE and ESE of the SEM, ESEM and RG-EPA algorithms
upon varying $N$ and $N_{B}$, and using the parameters in Table~\ref{tab:param}
with $M=2$, $K=10$, $\alpha=0.1$, $P_{max}^{B}=20$dBm, $P_{max}^{R}=10$dBm
and a cell radius of $0.75$km.}

\label{fig:multiplex} 
\end{figure}

Fig.~\ref{fig:multiplex} illustrates the effect of increasing $N$
and $N_{B}$ on the attainable SE and ESE. Note that in a similar
fashion to Fig.~\ref{fig:relays}, the SEM and ESEM algorithms attain
the same solutions in the operating region considered.

Observe from both Figs.~\ref{fig:multiplex}(a) and~\ref{fig:multiplex}(b)
that the attainable SE and ESE increase upon increasing
$N_{B}$. This is due to the increased attainable spatial degrees
of freedom at the BS in the first transmission phase, which allows
for more direct transmissions overall. However, both the SE and ESE
are reduced upon increasing $N$, which suggests that increasing the
number of subcarrier blocks does not increase the average efficiency
of each block. This is because the power constraints are fixed and
thus there is insufficient power for fully exploiting the additional
subcarrier blocks. However, note that both total SE and ESE do
indeed increase upon increasing $N$, which may be explicitly seen
upon multiplying the results of Figs.~\ref{fig:multiplex}(a) and~\ref{fig:multiplex}(b)
by $NW$. The RG-EPA algorithm performs worse in both cases as expected.

\section{Conclusions and future work\label{sec:Conc}}

In this paper, firstly a novel transmission protocol based on joint
transmit-BF and receive-BF was developed for the multi-relay MIMO-OFDMA
cellular network considered. This protocol allows for achieving high-SE
performance for the MIMO broadcast network consisting of a BS, multiple
RNs and multiple UEs. The associated MIMO channel matrices were mathematically
decomposed into multiple MISO channels, which we referred to as SMCs,
using receive-BF. By applying ZFBF at the transmitter, the
interference between SMC-based concurrent transmissions
is completely eliminated, provided that perfect CSI-knowledge is available.
For the purposes of obtaining a higher multiplexing gain, the SMCs
may be grouped according to the semi-orthogonality criterion. Consequently,
a pair of grouping algorithms were proposed, referred to as ESGA and
OCGA. The former exhaustively enumerates all of the possible groupings,
whereas the latter aims to be a lower-complexity design alternative.
Finding the SE-optimal and ESE-optimal SMC groupings as well as their
associated optimal power control variables were formulated
as optimization problems. With the aid of several variable relaxations
and transformations, these optimization problems were transformed
into concave optimization problems. Thus, the dual decomposition approach
was employed for finding the optimal solutions. We demonstrated that
the OCGA constitutes an attractive alternative to ESGA, since it offers
a near-optimal performance at a substantially reduced complexity.
Furthermore, several numerical results were presented for characterizing
the system's attainable SE and ESE performance across a wide range
of system parameters, such as the transmit power constraints, cell
radius, the number of RNs, the number of BS antennas and the number
of subcarrier blocks. Additionally, we demonstrated that our SEM/ESEM
algorithms perform significantly better than the benchmark RG-EPA
algorithm.

In our future work, we will consider unity frequency reuse multi-relay
multi-cell networks. Thus, these networks are interference-limited,
rather than noise-limited. Consequently, improved transmission protocols
and optimization methods are required for managing both the intra-cell
and inter-cell interference in order to improve the system's SE and
ESE performance.

\bibliographystyle{IEEEtran}
\bibliography{references}

\end{document}